\documentclass[aps,prl,singlecolumn,groupedaddress,longbibliography]{revtex4-1}
\usepackage{graphicx}
\usepackage{epstopdf}
\usepackage{dcolumn}
\usepackage{bm}
\usepackage{color}
\usepackage{amssymb, amsmath, bm, textcomp}

\usepackage[T1]{fontenc} 
\usepackage{float}       
\usepackage{mathrsfs}

\usepackage{setspace}    
\usepackage{color}
\usepackage{subfigure}

\begin{document}


\title{
Extended friction elucidates the breakdown of fast water transport in graphene oxide membranes}


\author{A. Montessori$^{1*\dagger}$, C.A. Amadei$^{2\dagger}$\\
\vspace{2mm}
G. Falcucci$^3$,M. Sega$^5$,C.D. Vecitis $^2$,and S. Succi $^{2,4 \dagger}$ }
\email[]{corresponding author: and.montessori@gmail.com}
\email[$^{\dagger}$]{These authors equally contributed to this work. A.M. and S.S. developed the mesoscopic Langevin approach. C.A.A. carried out all the experiments.
A.M., S.S. and C.A.A. wrote the manuscript.
C.V. supervised the experiments and helped in the writing of the manuscript.
G.F helped with the coding and contributed to the writing process. }%
\affiliation{$^1$ Dept. of Engineering - University of Rome ``Roma Tre'' \\
Via della Vasca Navale 79, 00141 Rome - Italy \\
$^2$ John A. Paulson School of Engineering and Applied Sciences, Harvard University,
Cambridge, MA 02138, USA\\
$^3$Dept. of Enterprise Engineering "Mario Lucertini" - University of Rome "Tor Vergata" - Via del Politecnico 1, 00133 Rome - Italy \\
$^4$ Istituto per le Applicazioni del Calcolo, CNR \\  
Via dei Taurini 19, 00185 Rome - Italy\\
$^5$ Institute of Computational Physics, University of Vienna, Sensengasse 8/9, 1090 Vienna, Austria
}


\date{\today}

\begin{abstract}

The understanding of water transport in graphene oxide (GO) membranes 
stands out as a major theoretical problem in graphene research. 
Notwithstanding the intense efforts devoted to the subject in the recent years, a 
consolidated picture  of water transport in GO membranes is yet to emerge.

By performing mesoscale simulations of water transport in ultrathin GO membranes, we
show that even small amounts of oxygen functionalities can lead to a dramatic drop of 
the GO permeability, in line with experimental findings.

The coexistence of bulk viscous dissipation and spatially 
extended molecular friction results in a major decrease of both slip 
and bulk flow,  thereby suppressing the fast water transport 
regime observed in pristine graphene nanochannels.

Inspection of the flow structure reveals an inverted curvature in the near-wall region,
which connects smoothly with a parabolic profile in the bulk region. 
Such inverted curvature is a distinctive signature of the coexistence between 
single-particle Langevin friction and collective hydrodynamics. 

The present mesoscopic model with spatially extended friction may offer a 
computationally  efficient tool for future simulations of water transport in nanomaterials. 

\end{abstract}


\maketitle

Water transport through graphene-derived membranes has recently captured 
much attention due to its promising potential 
for many technological applications, such as water filtration, 
separation processes, and heterogeneous catalysis 
\cite{cohen2012water,mishra2011functionalized,julkapli2015graphene,garberoglio2007inhomogeneity}. 
A number of simulations have shown that fast water permeation through carbon materials, such as carbon nanotubes (CNT \cite{joseph2008carbon,holt2006fast,zuo2009transport,majumder2005nanoscale}) and pristine graphene membranes \cite{nair2012unimpeded} is due to the slip flow at the water-carbon interface. Fast water transport (FWT) with permeabilities higher than $10 \; L m^{-2} h^{-1} bar^{-1}$ \cite{hu2013enabling,jiang2015engineered,han2013ultrathin} has also
been reported for graphene oxide laminate (GOL). \\
However, in contrast to pristine graphene and CNT, a clear consensus on the GOL fast water transport mechanism has yet to emerge \cite{nair2012unimpeded,huang2013ultrafast,sun2016recent}. To date, departures from hydrodynamic (i.e., Hagen-Poiseuille) behavior are typically attributed to the low friction experienced by water in atomistically smooth graphene nanochannels \cite{joshi2014precise} or to  the presence of structural defects in GOL, leading 
to shorter water paths \cite{wei2014breakdown,joseph2008carbon}. 
The latter hypothesis is further corroborated by the breakdown of FWT from molecular interactions of water with basal plane 
hydroxide and epoxide groups, which hinders the motion of water molecules. This phenomenon was recently reported in non-equilibrium molecular dynamics simulations of 
flows inside  graphene oxide (GO) nanochannels, \cite{wei2014breakdown,wei2014understanding}. 
Under such conditions, slip-corrected continuum hydrodynamics 
is expected to provide a correct description of the aforementioned
transport mechanism. However, at the hydrodynamic level
the details of the slip corrections must be imposed a priori because
the physics of slip flow is governed by
non-equilibrium phenomena. 
Such details can certainly be accounted for by a fully atomistic 
approach, but at the price of prohibitive computational costs.
All of the above lead to an ideal scenario for {\it mesoscale} 
modeling techniques, which may offer a valuable compromise between 
physical fidelity and computational viability, \cite{montessori2015lattice}.\\
In this work, we explore water permeation in GOL via mesoscale simulations.
Furthermore, we provide a direct calculation of the membrane's
permeability, which is drastically lower than the one observed in pristine graphene characterized by FWT.
Our simulations exhibit satisfactory agreement with experimental GOL permeability and also provide values of the slip length 
on par with recent non-equilibrium molecular dynamics 
simulations (NEMD) \cite{wei2014understanding}.
More precisely, we demonstrate that a suitably simplified (lattice)
kinetic model proves capable of predicting the breakdown of FWT, namely the 
macroscopic permeability of the GO membrane, as well as the slip phenomena, at a 
very affordable computational cost. For instance, a baseline simulation takes only
a few CPU hours on a high-end PC.
\begin{figure*}
\centering
\includegraphics[scale=0.50]{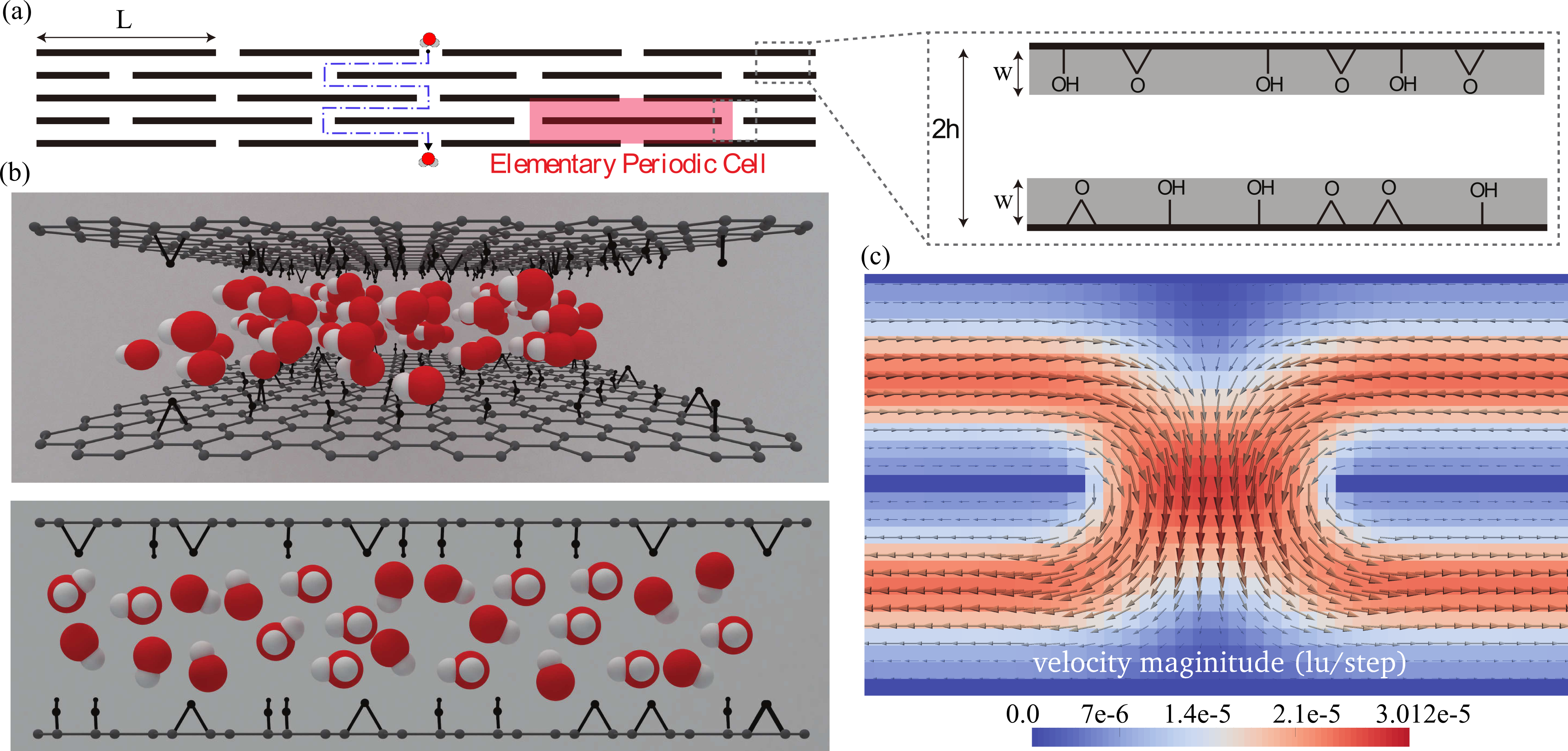}
\caption{\label{hydroxyl}  Sketch of the GOL structure with a zoom of the GO nanochannel decorated with oxygen functionalities (panel (a)). 
In the sketch, L is the GO flake's length, $2h$ is the spacing between two GO layers and w is the spatial extent of the Langevin-like frictional force.
The red area in the GOL structure identifies the elementary periodic
cell used in the simulation.
As shown in panel (b), hydroxide abd epoxide groups  interact with the water molecules
slowing down their motion inside the GO nanochannels. 
In panel (c), the water molecules flow from the inlet port (top) to the
outlet port (bottom), under the effect of applied pressure.
The vertical motion is hindered by a series of horizontal staggered plates
(GO flakes), which force the water molecules to follow a tortuous path 
from inlet to outlet ports. 
}
\end{figure*}

\textbf{Method.}
GO was synthesized by modified Hummers' method \cite{kovtyukhova1999layer} and dispersed in ethanol at a concentration of $0.1\;mg/L$. Additional details can be found in the supporting information. The solution was then bath sonicated for 5 minute in a Branson 2510 Ultrasonic cleaner. $1\;ml$ of the sonicated GO solution was deposited on a porous polyvinylidene difluoride (PVDF) membrane (pore size = $200\;nm$) via vacuum filtration. The membranes were then characterized by using a Zeiss ULTRA Field Emission Scanning Electron Microscope (SEM) with an In-lens detector (see Figure S\;1 and S\;2 in the supporting information). Scanning electron microscopy and mass analysis suggest the formation of a circa $30\;nm$ thick GO layer on top of the PVDF membrane. The crystallographic structure of the membranes was analyzed with a Bruker D8 equipped with a two-dimensional VANTEC-500 detector. The spectra were obtained by the integration of the 2D diffraction pattern image (see Figure S\;3 in the supporting information) via EVA software. Depending on the sample, the integration time was between 600-1200 seconds. Permeability results were carried out with a custom-made dead-end filtration system operating at a maximum pressure of $3.5\;bar$. In particular, a $2\;cm$ in diameter GO membrane was cut and placed in a stainless steel EDM Millipore filter holder. The pressure was monitored with an Ingersoll pressure gauge regulator (see Figure S\;4 in the supporting information).
The numerical simulations are based on the Lattice Boltzmann (LB) method, augmented 
with a novel Langevin-like frictional force accounting for the GO-water
interactions. 
Since the LB method is largely documented in the literature 
\cite{qian1992lattice,higuera1989lattice,succi2038lattice}, in the 
following we discuss only its basic features.
The lattice Boltzmann equation reads as follows:
\begin{equation}
f_i(\vec{x}+\vec{c}_i \Delta t,t+\Delta t) = (1-1/\tau) f_i + 1/\tau f_i^e + \Delta f_i
\end{equation}
where $f_i(\vec{x},t)$ is a set of discrete probability distribution functions
(PDFs) representing the probability of finding a molecule at 
position $\vec{x}$ and time $t$ with a lattice-constrained velocity $\vec{c}_i$, 
where the index $i$ runs over the nine directions of the lattice \cite{succi2001lattice}.
$f_i^e$ is the set of discrete  local Maxwellian equilibria (i.e., truncated low-Mach number expansion of the Maxwell-Boltzmann distribution)
 \cite{qian1992lattice} and $c_s$ is the speed of sound of the model \cite{succi2001lattice,wolf2000lattice}.
In the above equation, the left hand side is the lattice transcription of molecular free-flight 
along the lattice directions, while the right hand side describes collisional
relaxation towards local equilibrium, described by a low-Mach number expansion 
of the Maxwell-Boltzmann distribution.
The relaxation parameter $\tau$ controls the kinematic viscosity of the lattice fluid through the relation $\nu = c_s^2(\tau-1/2)$ in lattice units $\Delta x = \Delta t =1$.
The last term, $\Delta f_i$, is the correction due to the friction
exerted by the hydroxyl and epoxy (see Fig. \ref{hydroxyl}b) on the water molecules and can be expressed as: $\Delta f_i = w_i \frac{\vec{F} \cdot \vec{c_i}}{c_s^2}$,
in which $w_i$ is the set of weights for the chosen lattice \cite{qian1992lattice},
and
the frictional force is taken in the following Langevin form \cite{smiatek2008tunable}: 
\begin{equation}
\vec{F} = -\rho \gamma(y) \vec{u}
\end{equation}
where $\rho$ and $\vec{u}$ are the fluid density and velocity respectively and the heterogeneous friction coefficient reads as follows:
\begin{equation}
\gamma(y) = \gamma_0 
(e^{-\frac{y}{\text{w}}} H_L(y,\delta) + e^{\frac{(y-h)}{\text{w}}} H_R(y,\delta))
\end{equation}
where w is a representative size of the protruding functional groups and
$\gamma_0$ is a characteristic water-hydroxyl collision frequency, taken equal to $ 0.2 $ (in lattice units) in all the 
simulations, corresponding to a collision rate of about $70 \; ps^{-1}$ \cite{pastor1988analysis,izaguirre2001langevin}.
The wall function $H_L(y)$ takes the value $1$ for $0<y<\delta$ and $0$ elsewhere.
Likewise, $H_R(y)=1$ for $2h-\delta<y<2h$ and $0$ elsewhere.
The reference case is $\delta=\text{w}$ (truncated Langevin throughout the text).
In this case, if $\text{w}<h$, the Langevin force drops discontinuously 
to zero in the central region of the channel, w$< y < 2h-$w.
To regularise this discontinuity, we also consider the case $\delta=h$ (non-truncated Langevin).
The physical idea behind the Langevin-like frictional force
is to account for the complex water-GO molecular interactions
at a coarse-grained level, whereby all atomistic details are 
channeled into the parameters $\gamma_0$ and w.

Whether the contact angle and the slip length can be treated as independent 
variables, still is an open question in the current literature (see \cite{bocquet2008prl,sega2013regularization}).
In line with the mesoscale nature of our model, we assume sufficient universality to support 
a direct correlation between the water-graphene contact angle and the slip length. 
This said, from the NEMD results (Fig. 4d in  \cite{huang2013ultrafast}), we read off a slip length 
between $0.3-1.0 nm$ with a $40$ $\%$ to 5$\%$ of hydroxyl groups, respectively.
According to the authors, these values correspond to a contact angle $\theta=29.1^{\circ}$, with 20$\%$ 
hydroxyl groups and  a $50\;nm$ slip length in pristine graphene. 
Based on these data, it is reasonable to assume that the slip length should be of the order of the molecular 
size of the hydroxyl groups ($0.1-0.2\;nm$), which is precisely the assumption made in our model.  
This phenomenological approach allows us to import knowledge at the microscale within a mesoscale computational harness. In particular, the model permits us to reach space-time scales of experimental relevance 
without being tied-down to the continuum assumptions
The lattice units are $\Delta x = 10^{-10}\;m$, yielding a time step $\Delta t \sim  3 \cdot 10^{-15}\;s$. 
Note that sub-molecular spatial resolutions are typical of
LB simulations of nanoflows \cite{horbach2006lattice}.
The time-step, however is about an order of magnitude
larger then the time steps typically used in MD simulations of flows through GO interlayers \cite{huang2013ultrafast}.
It is also worth mentioning that  the CPU time needed to update a single molecule is 
significantly larger than the one required to advance a single LB cell, because the cell contains 
less neighbours and, more importantly, such neighbours are fixed in space, hence there is
no need to recompute the interaction list at every time-step.
\cite{fyta2006multiscale}. Moreover, LB requires no statistical averaging, since it is 
based on a pre-averaged probability distribution function. 
In addition, LB is often more efficient than computational fluid dynamics 
because i) pressure is available locally in space and time, with no need of solving
a demanding Poisson problem, ii) transport is exact, since free-streaming proceeds
along fixed molecular velocities instead of space-time dependent material 
streamlines, iii)  diffusion is emergent, hence it does not require second order spatial 
derivatives, thus facilitating they formulation of boundary conditions in confined flows \cite{succi2038lattice}.


\textbf{Results.}
Assuming the GOL structure to be symmetric and periodic \cite{han2013ultrathin,nair2012unimpeded} (see red area in Fig. \ref{hydroxyl}a), we consider only 
two nanochannels out of the full device. A similar geometry set-up has been recently employed to investigate water permeation through graphene-based membranes by means of MD simulations \cite{muscatello2016opt}. 
These two channels are
connected via two openings of half the width of the inlet/outlet pores (see Fig. \ref{hydroxyl}a).
The boundary conditions at the left-right and top-bottom surfaces are 
periodic, to simulate the proximity of two adjacent GO layers.
At solid walls, the molecules experience the Langevin friction previously described.
\begin{figure}
\includegraphics[scale=0.3]{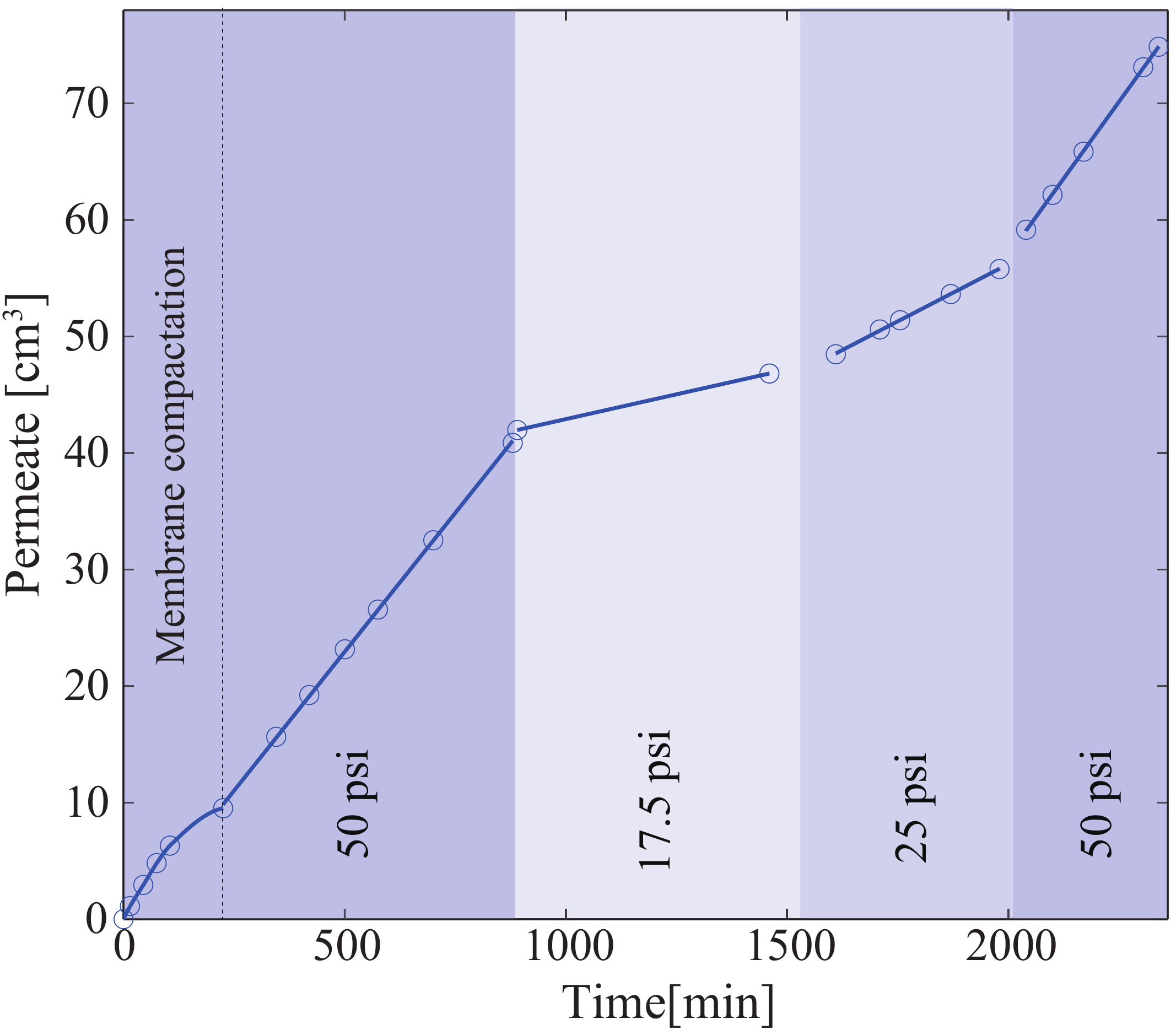}
\caption{Water permeate versus time obtained for  
a GO membrane with a flake's length $L\sim1 \; \mu m$ and channel spacing 
$2h\sim0.8\;nm$.  
The permeate increases linearly with time, thus denoting a hydrodynamic (Darcy-like) behaviour of the membrane. See also the flow rate vs the pressure gradient plot reported in figure S\;5 of the supplementary information, reporting a linear behaviour typical of the hydrodynamic regime.
Dashed line indicates the compaction time of the membrane.  
}\label{permeate}
\end{figure}
We have run several simulations with different values of w.
The grid resolution is  $10000 \times 20$,
corresponding to a flake length of $10^{-6}\;m$ , which we take
as a representative experimental value, \cite{amadei2016fabrication}. The mesoscale model was tested against 
experimental measurements of permeability.
The permeability of the membrane is defined as 
$\kappa=\mu u_s/ |\nabla p|$, where $\mu$ is the dynamic viscosity, $\nabla p$ 
the pressure gradient across the membrane and $u_s$ the superficial 
velocity defined as the ratio between the membrane discharge per unit area.  
It is worth recalling that in the Darcy regime, the permeability 
$\kappa$ is pressure independent (see figure S\;5 in the supporting information).
In order to test our numerical results against experimental data, we inferred the 
experimental values of flow rates (mass flow per unit time) from the permeate vs 
time plot, reported in Fig. \ref{permeate}, for different values of the 
applied pressure. 
From these data, we compute a dimensionless permeability $\kappa^*=\kappa/(2h)^2$,
where $2h$  represents the spacing between two GO layers.
X-ray diffraction measurements give $2h\sim\;0.8nm$.
The experimental value of the dimensionless permeability $\kappa^*$ for a $\sim30 \; nm$ 
thick membrane of diameter $d=2 \; cm$ is $\kappa^*\sim 2.8 \times 10^{-4}$, corresponding to a permeability of $3.6 \pm 0.5\; L MH/bar$.
This value of permeability is in line with data previously reported for ultrathin ($< 50\;nm$) GO membranes ( see   \cite{han2013ultrathin,kovtyukhova1999layer,hu2013enabling}) and represents a promising result for nanofiltration applications. It is important to underline that higher values of permeability have been achieved for GO membranes. However, these values can be connected to the presence of defects or larger GO nanochannels due to  the chemical modification of the GO membranes \cite{han2013ultrathin,xia2015ultrathin,ying2014plane}. Higher permeability can also be achieved by intercalating the GO membranes with high aspect ratio nanoarchitectures, such as CNT \cite{huang2013ultrafast,goh2015all}.

\begin{figure*}
\subfigure{\includegraphics[scale=0.4]{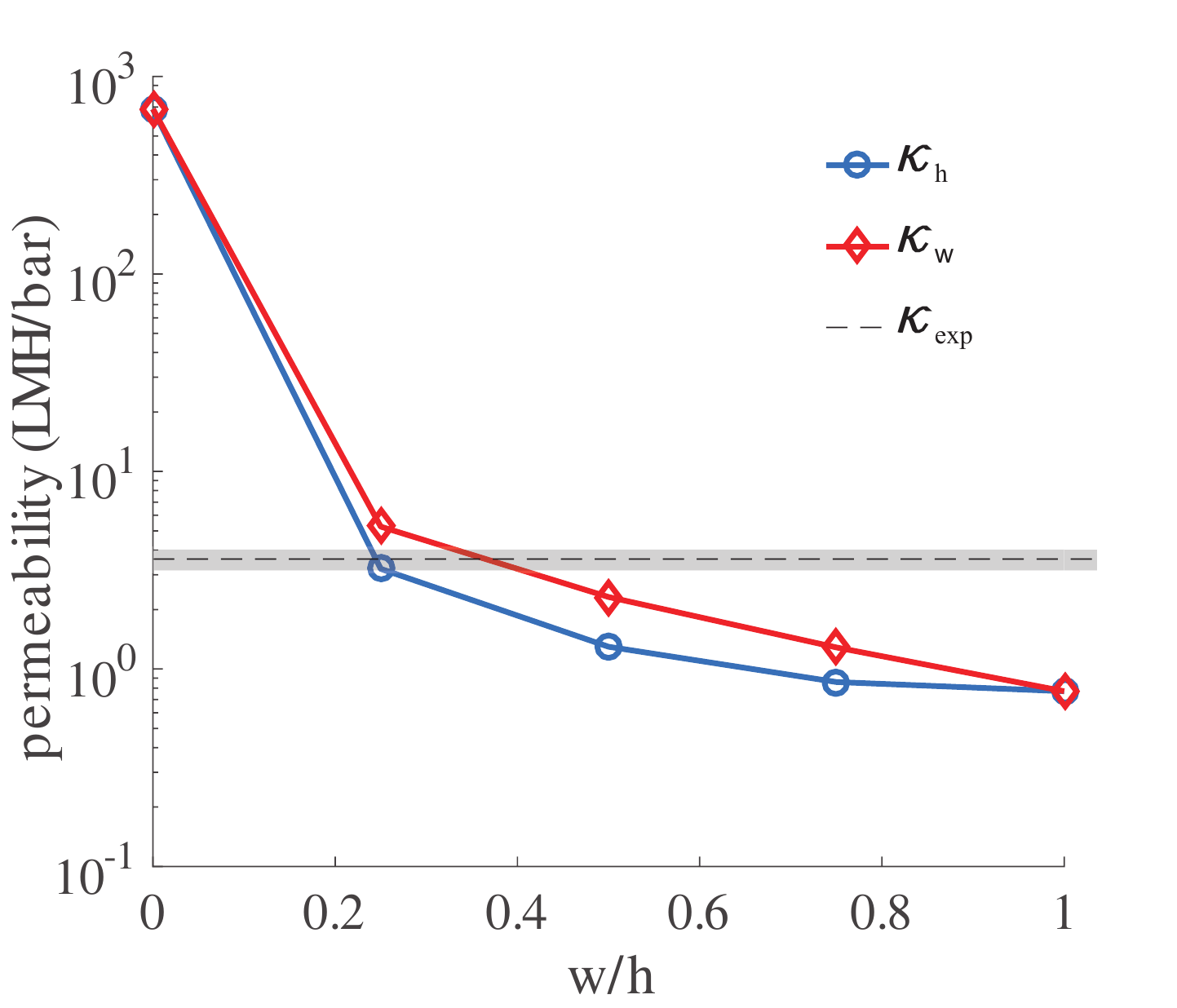}}\
\subfigure{\includegraphics[scale=0.4]{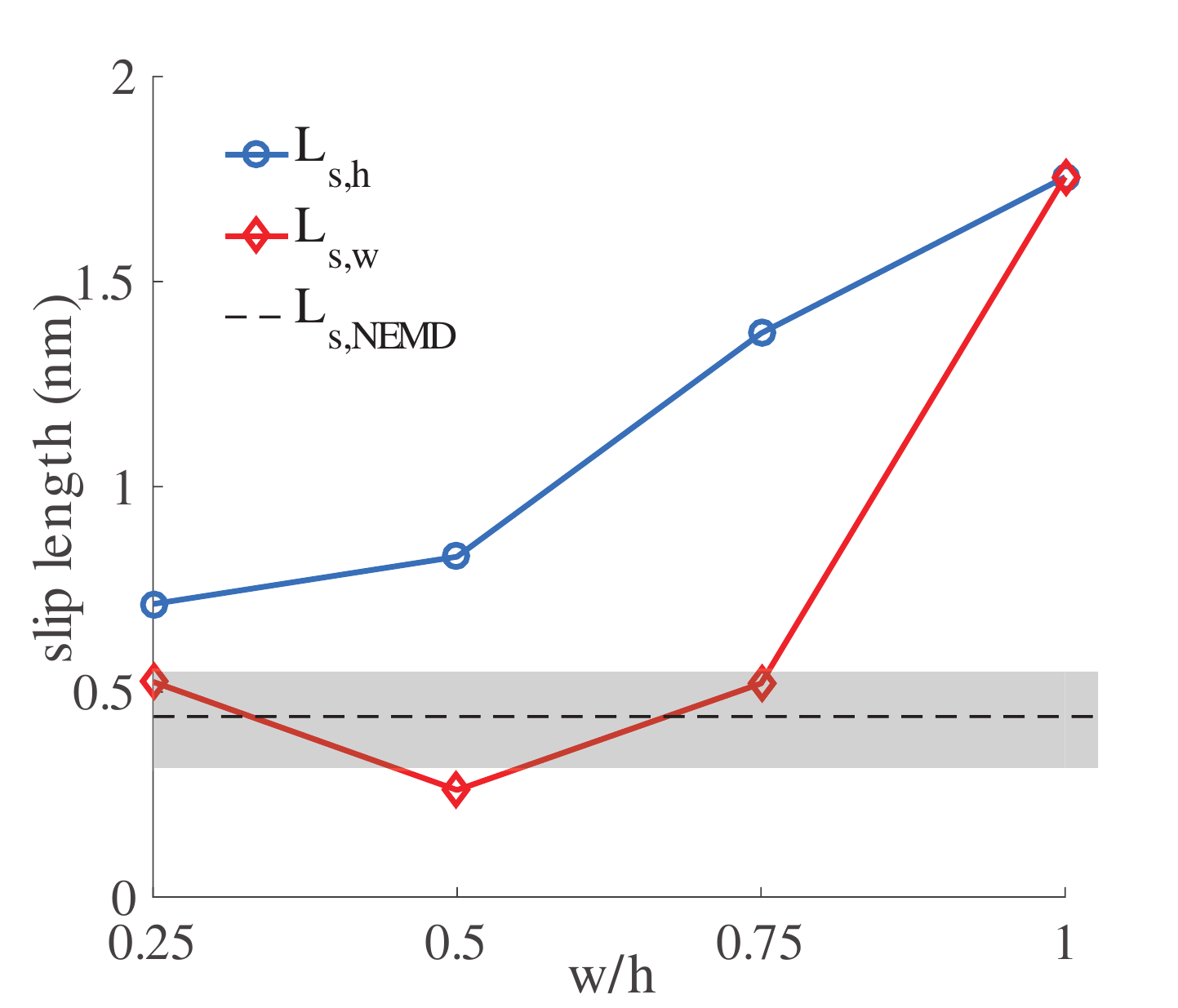}}
\caption{\label{kappa} Left panel: Solid lines with symbols represent the simulated permeabilities (truncated ($\kappa_h$) and  non-truncated ($\kappa_\text{w}$)) while black dashed line reports the average experimental ($\kappa_{exp}$) value as a function of  w$=0.1$, $0.2$, $0.3$, $0.4\;nm$ and $h=0.4\;nm$. 
Right panel: solid lines with symbols represent the slip length computed with the LB approach  (truncated ($L_{s,w}$)   and non-truncated ($L_{s,h}$)) while the dashed line stands for the NEMD average slip length \cite{wei2014breakdown}.
In both figures grey shaded area identifies the the error bars of the experimental permeability and of the NEMD slip lengths.
The case w$=0$ correspond to free-slip conditions, taken as representative of the FWT consitions (see the main text). 
}
\end{figure*}
Fig. \ref{kappa} illustrates two comparisons, for both the truncated and non-truncated Langevin frictions: i)  between the  permeability obtained by mesoscopic simulations and experiments; ii) between the slip length obtained by NEMD \cite{wei2014breakdown} and by this study. The case w$=0$ corresponds to a simulation without the Langevin friction, i.e. pure {\it free-slip} hydrodynamics.
By free-slip we mean boundary conditions which leave the flow momentum
tangential to the wall unchanged. 
This is representative of the FWT regime observed in pristine
graphene experiments. 
Note that the flow still reaches a steady-state solution on account of the 
localized dissipation experienced at the sharp $90^\circ$ turns between 
two subsequent layers, visible in Fig. \ref{hydroxyl}c.
The main outcome from Fig. \ref{kappa} is a dramatic drop in
permeability of two orders of magnitude,
already at w$=0.1 \; nm$, i.e. w$/h=0.25$.
Such a dramatic drop in permeability is consistent with 
experimental observations that report a suppression of FWT regimes
in the presence of hydroxyl groups. 
Further increments of w lead to a sizable
reduction in permeability by circa one order of magnitude, from 
w$=0.1\;nm$ to w$=0.4\;nm$. It is worth noting that, the employed resolution ($\Delta x=0.1 \; nm$) poses a constraint on the lower bound of w.
We note that the truncated ($\kappa_\text{w}$, $\delta=\text{w}$) and non-truncated 
($\kappa_h$, $\delta=h$) scenarios yield nearly the same picture with only minor 
quantitative variations.
Moreover, Fig. \ref{kappa} shows that for both the truncated and non-truncated scenario, the slip length remains
between $0.5 \;nm$ and $2\;nm$,  in agreement with the values provided by NEMD \cite{wei2014breakdown}. 
It is worth noting that the best match of the experimental (permeability) and NEMD (slip lengths) results with the simulations  is obtained when w is between $0.1\;nm$ and $0.2\; nm$ (i.e. w$/h=0.25 \div 0.5$), which agrees with the physical dimension of the oxygen functionalities in the GO nanochannels \cite{haynes2014crc}. 
This further corroborates the validity of the model and its potential to capture the physical phenomena in 2D nanostructured  inspired 
materials within an efficient computational framework.

\textbf{Inside the flow structure.}
Next, we proceed  to inspect the internal structure of the flow.
In particular, we focus on the occurrence of slip flow in the presence
of hydroxyl, as recently observed in non-equilibrium molecular 
dynamics simulations \cite{wei2014breakdown}.
As previously discussed, slip flow is a typical manifestation of 
individual non-hydrodynamic behavior driven by non-equilibrium 
effects near the wall.
To glean quantitative insights into such non-equilibrium phenomena  is informative for inspecting the one-dimensional cross-flow
profiles $u_x(y)$ for different values of w versus the
case of free-slip boundary conditions.
\begin{figure}
\subfigure[]{\includegraphics[scale=0.43]{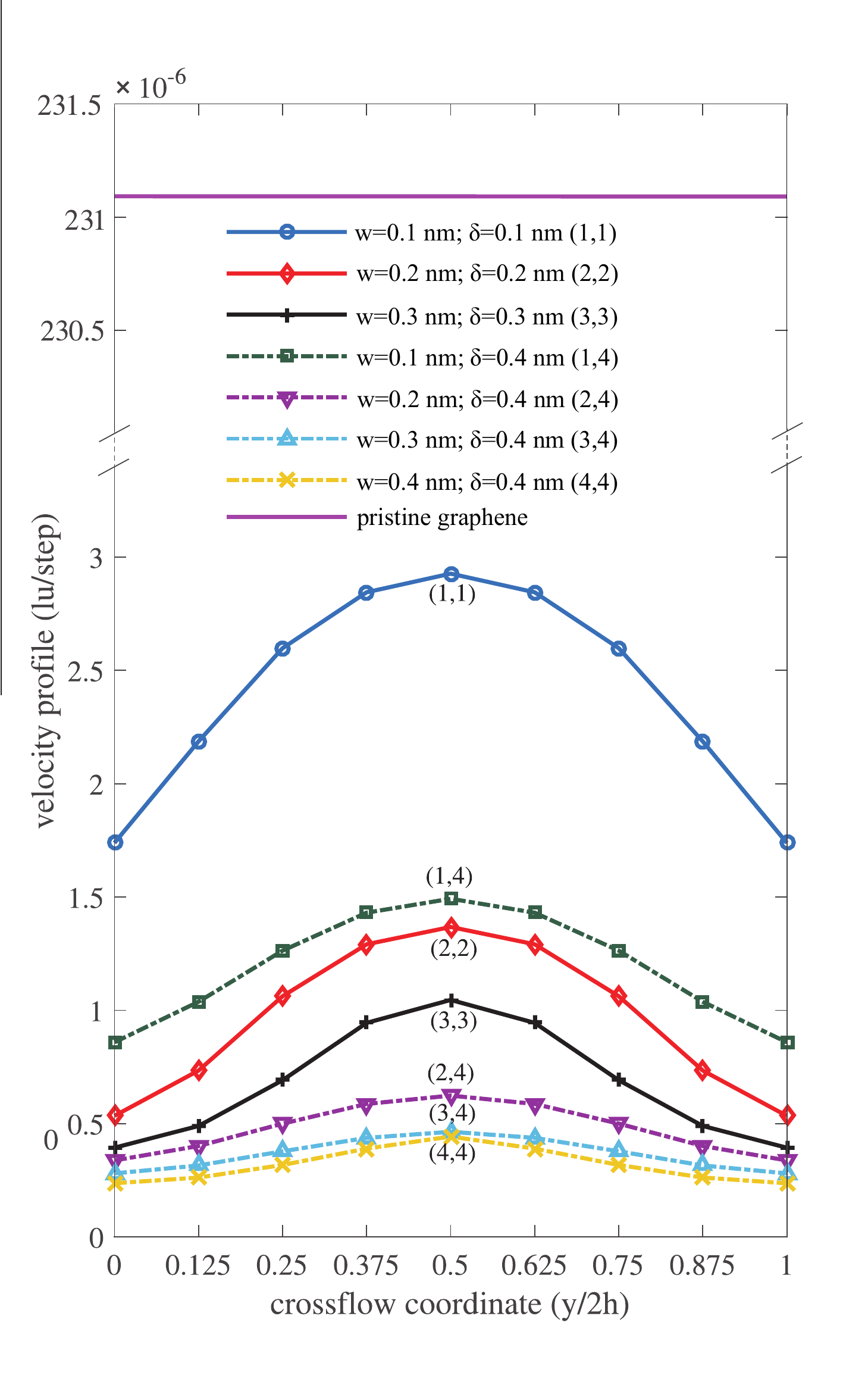}}
\subfigure[]{\includegraphics[scale=0.5]{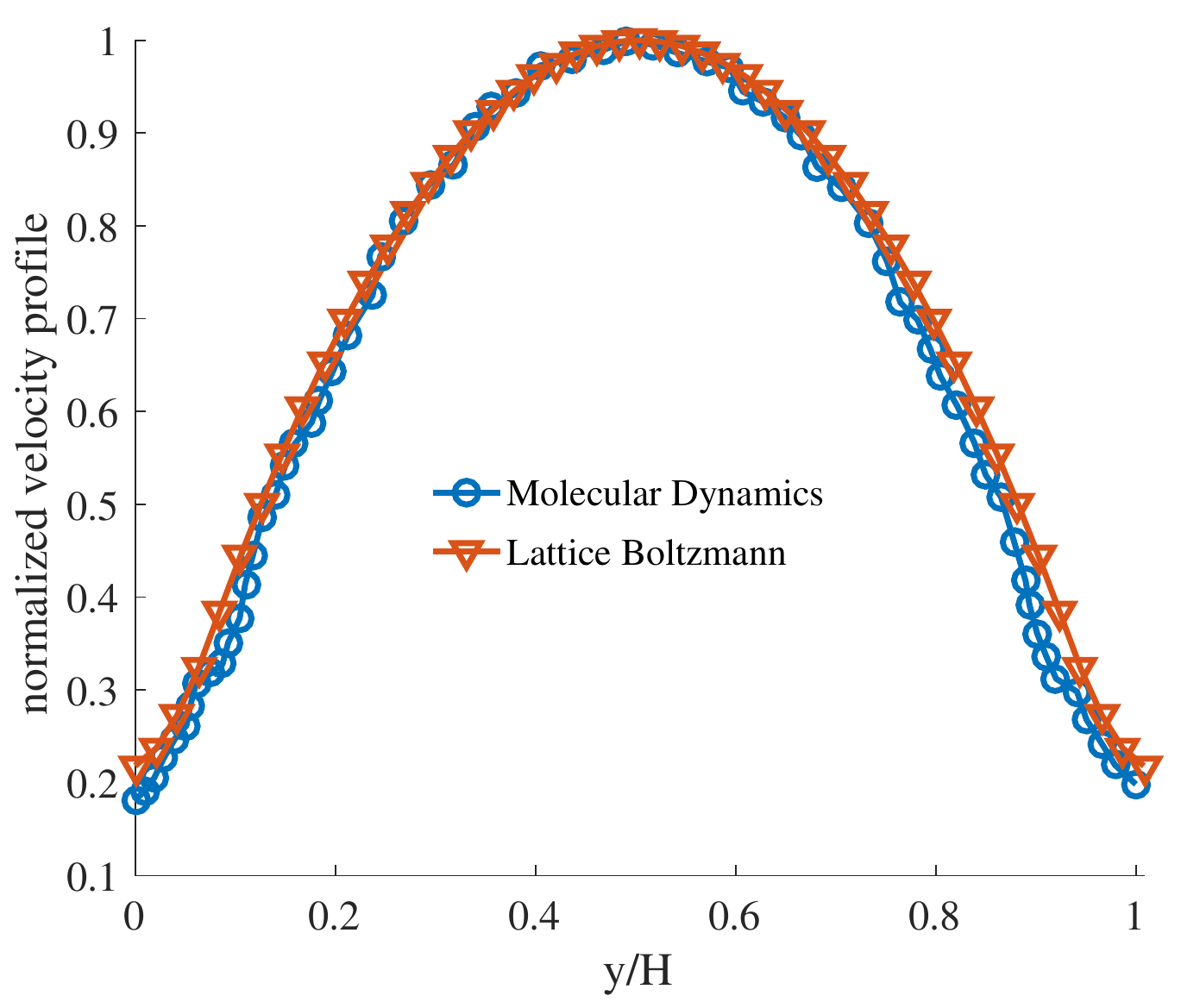}}
\caption{(a) Flow profiles $u_x(y)$ for different values of the friction 
length w and cutoff length $\delta$, $(\text{w},\delta)$. The horizontal line refers to
the free-slip flow in the absence of Langevin friction. 
The two numbers within parenthesis denote the values of w and $\delta$.
Friction and cutoff lengths are made dimensionless by dividing them by  half of the channel spacing $h=0.4\;nm$.
In panel (b) we report the velocity profile obtained by the Langevin-LB on a $3\;nm$ wide channel flow using $50$ lattice point compared to the MD profile taken from \cite{huang2013ultrafast}.On
the $x$-axis the non-dimensional channel width ($\text{y}/2h$) is reported.
The profiles are rescaled by the peak value of the
velocity.
}
\label{flowprofiles}
\end{figure}
Fig. \ref{flowprofiles}a illustrates that the flow profiles
display a Poiseuille-like parabolic trend in the bulk region, smoothly 
turning into a flat profile near the wall, with a positive 
curvature and a small non-zero slip velocity.
The comparison between  the pristine graphene profile with the other profiles  highlights the major drop of mass
flow due to Langevin friction. 
The increase of  w leads to the suppression and flattening of the water profiles.
From the velocity profiles, we compute the slip length  according to
$L_s = \lim_{y \to 0} \; |\frac{u_x(y)}{\partial_y u_x(y)}|$.
As previously discussed for Fig. \ref{kappa}, we confirm the presence of a residual slip length, which is small in 
absolute physical units, but fairly sizable in units of
the channel width $h$, namely $L_s/2h \sim 0.5$. Furthermore, the effect of the different cut-off lengths employed in the truncated $(\delta=\text{w})$ and non-truncated $(\delta=h)$ Langevin is minor, leading to slight differences in both the membrane permeability and slip lengths.
To further test the robustness of this approach in Fig. \ref{flowprofiles}(b) we report a comparison between the velocity profiles obtained by the Langevin-LB and the MD approach \cite{huang2013ultrafast} for a $3\;nm$ wide GO channel, showing an outstanding agreement between the two models. 

We conclude this Letter with an ex-post analysis of our mesoscale
model in light of the numerical results discussed above. First, we note that although both friction and viscous dissipation withstand the driving action of the pressure gradient, they 
operate according to very different and competing mechanisms.
Friction drives the fluid towards the following local flow configuration:
\begin{equation}
\label{UL}
u_x^{(L)}(y) = \frac{|\nabla_x p|}{\rho \gamma(y)} 
\end{equation}
This results in a slip-flow $u_{slip}=|\nabla_x p|/(\rho \gamma_0)$ (independent of w) 
at the wall and the corresponding slip length  is $L_s=$w. 
Viscous dissipation, on the other hand, drives the fluid towards a 
parabolic Hagen-Poiseuille profile, 
\begin{equation}
\label{UH}
u_x^{(H)}(y) = u_H [ \;\frac{y}{h}(2-\frac{y}{h}) + Const.].
\end{equation}
where we have set $u_H = \frac{\nabla_x p }{\rho \nu / h^2}$.
The above hydrodynamic solution is compatible with either slip or non-slip flow 
conditions, depending on the value of the constant at the right hand side.  
In slip-hydrodynamics $Const.=L_s/h$, so that $L_s/h \to 0$ recovers standard no-slip hydrodynamics.  
This value can only be prescribed a-priori, exposing the weakness of the hydrodynamic approach previously mentioned.

\begin{figure}
\includegraphics[trim={3cm 0 0 0},scale=0.5]{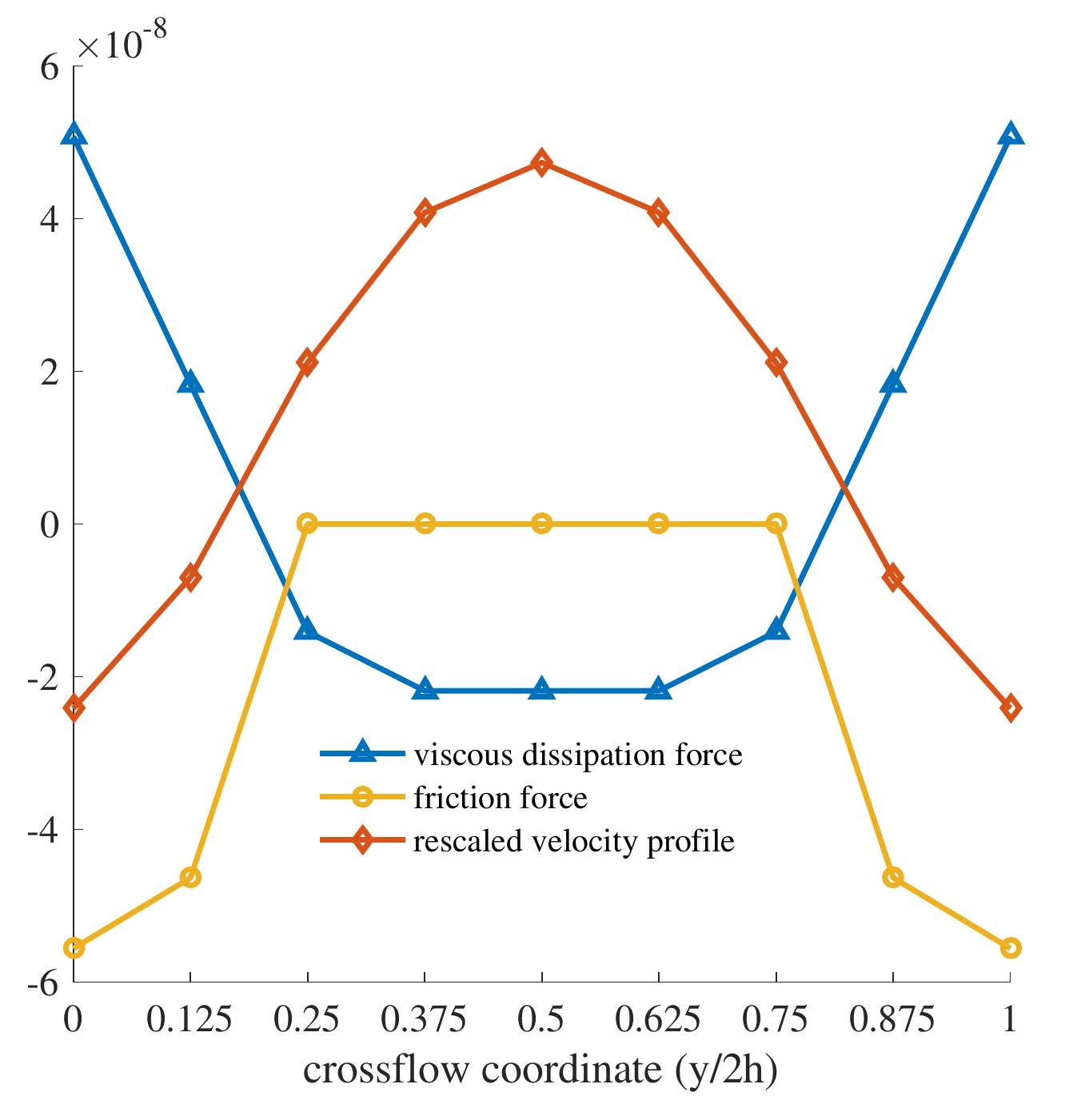}
\caption{Friction and viscous dissipation forces (in lattice units) as a function of 
the crossflow coordinate $y/2h$ for the case w$=0.2\;nm$, $\delta=0.2\;nm$.
For the sake of reference, the corresponding (rescaled) flow profile 
is also reported.
\label{figforce}
}
\end{figure}

The two profiles, $u_x^{(L)}(y)$ and $u_x^{(H)}(y)$ cannot coexist other than in an intermediate 
compromising form, resulting from the smoothing of the Langevin
profile due to viscous dissipation. To gain a deeper understanding, in Fig. \ref{figforce} we report 
the friction and viscous forces separately, for the case w$=\delta=0.2\;nm$.
As one can see, the bulk flow is dominated by viscous dissipation (friction is 
zero in the bulk because $\delta=0.2\;nm$), while in the vicinity of the wall, friction takes over. However, the
two contributions become comparable but opposite in sign due to the positive curvature
of the flow profile. We have also checked through separate simulations that the sign inversion of hydrodynamic
dissipation holds true at higher resolution (see Fig. \ref{flowprofiles}(b)).
The inversion of the dissipative force, namely $F_d = \rho \nu \frac{\partial^2 u}{\partial y^2}>0$ at the wall 
is a genuine effect of the spatially distributed decaying Langevin friction. 
Indeed, since the free-slip boundary condition forces $\nu \frac{\partial^2 u}{\partial y^2}=0$, the 
near-wall dissipative force is  dominated by  the Langevin friction, which is incompatible with $F_d=0$. 
Since Langevin friction decays exponentially away from the  
wall, the near-wall flow profile  must grow exponentially according to eq.\ref{UL},  
thus exhibiting the observed inverted (positive) curvature. 
Because of the exponential decay of the extended friction, the bulk region  is still 
dominated by viscous dissipation, as reflected by the bulk parabolic profile clearly visible in fig.\ref{figforce}.  
The onset of the inverted curvature is a distinctive signature of the
coexistence between single particle Langevin friction (slip-flow) and collective hydrodynamics (bulk  flow). 
Differently from continuum methods, which impose the slip length 
through the boundary conditions,  in our model it naturally emerges
from the self-consistent competition between  extended friction and hydrodynamic dissipation.
This extra-freedom is key to recover the inverted curvature profile.

\textbf{Conclusions.}
In summary, our numerical simulations portray the following picture.
Even a small amount of spatially extended Langevin friction, w$/h=0.25$, leads to a dramatic
drop in the mass flow compared to free-slip hydrodynamics. 
Such friction still supports a small residual slip-flow at the wall, with 
a slip length of the order of the friction length w. However, this flow 
is largely negligible compared to the free-slip in the absence of
Langevin friction. The net result is a dramatic loss of permeability due to 
the presence of the functional groups, hence the inhibition of 
the FWT regime observed in pristine graphene membranes. 
Viscous effects dominate the bulk flow and contribute to the smoothing of
the sharp features of the flow due to the presence of the hydroxyl and epoxy.
Inspection of the flow structure reveals an inverted curvature in the near-wall region,
which connects smoothly with a parabolic profile in the bulk region 
Such inverted curvature is a distinctive signature of the coexistence between 
single particle Langevin friction and collective hydrodynamics. 

\section{Supporting Information} 
\subsection{Hummers' Method}
\begin{figure}
\includegraphics[scale=0.4]{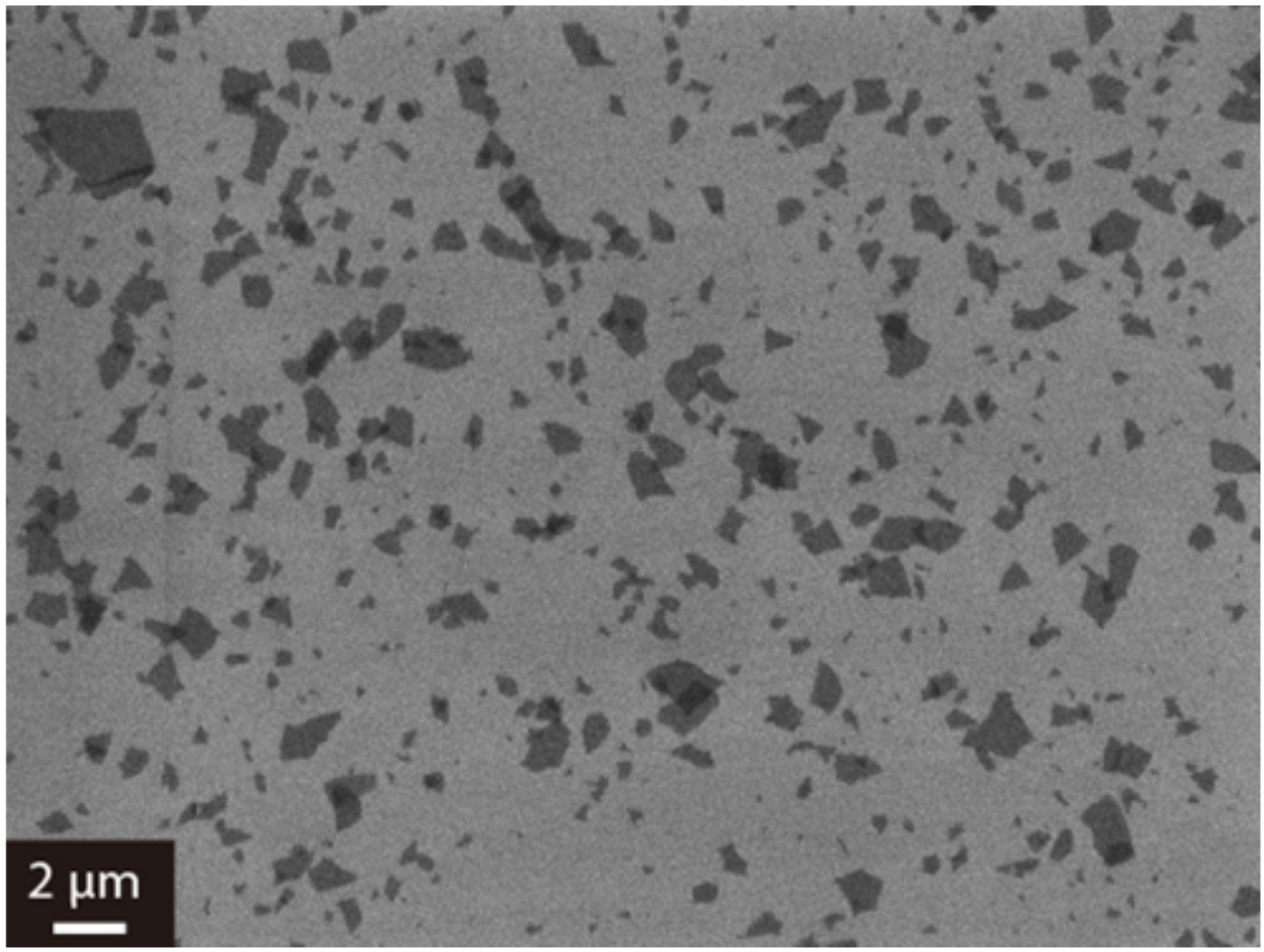}
\caption{\label{S1} monolayer GO flake's on a silica wafer.}
\end{figure}

GO solution was prepared using a modified Hummers' method with additional pre-oxidation of the graphite flakes. $5 g$ of the graphite powder were pre-oxidized using sulfuric acid ($30 mL$; $97$$\%$ $H_2SO_4$), phosphorus pentoxide ($4.2 g$; $P_2O_5$), and potassium persulfate ($4.2 g$; $K_2S_2O_8$) in a water bath at $75^{\circ} C$ for $4.5 h$. The mixture was then cooled to room temperature and diluted with $700 mL$ of deionized water (DI) and vacuum filtered through a poly(tetrafluoroethylene) membrane (pore size $5 \mu m$). After pre-oxidation, the modified Hummers' method was carried out by suspending the pre-oxidized material in $H_2SO_4$ ($150 mL$; $97$$\%$) in an ice bath for $20 min$. 
Potassium permanganate was slowly added ($15 g$; $KMnO_4$) and the solution was heated to $35\circ C$ for $2 h$. The process was completed by adding $250 mL$ of DI water and heating the mixture to $70^{\circ} C$ for an additional $2 h$. The reaction was quenched with hydrogen peroxide ($30 mL$; $H_2O_2$) and DI water ($750 mL$). In order to quench the unreacted reagent and clean the solution, the product was then filtered through a $300 \mu m$ testing sieve, then through glass fiber and centrifuged for $1 h$ at $5000 RPM$ (Sorvall RC-5C plus). The supernatant was discared whereas the precipitate was washed with hydrochloric acid ($400 mL$, $10$$\%$ $HCl$). The sieving, centrifugation, and washing process was then repeated using DI (two times) and ethanol ($EtOH$, two times). The final solution was then dispersed in $300 mL$ of $EtOH$ at a concentration of $\simeq 0.1 mg/L$. The overall yield of the process was circa $5$$\%$. Figure S\;\ref{S1} represents a SEM image of a drop casted GO solution on a silicon wafer. As one can see, more than $90$ $\%$ of the GO flakes is monolayer with an average area of $1.5 \pm 0.4 \mu m^2$.

\subsection{Membrane characterization}
\begin{figure}
\includegraphics[scale=0.4]{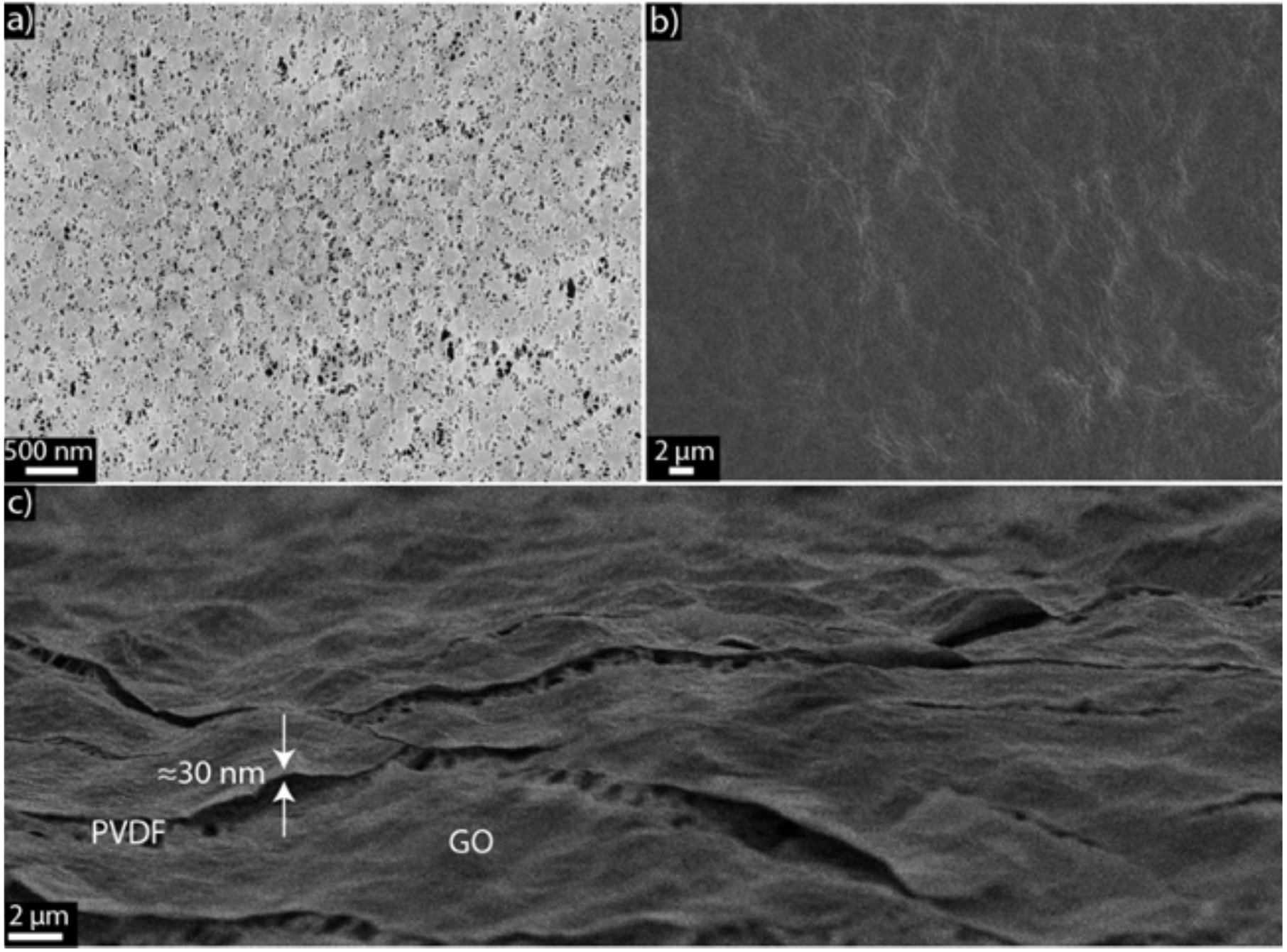}
\caption{  SEM characterization of (a) bare PVDF membrane, (b) GO membrane on PVDF and (c) cross section of GO membrane.}\label{S2}
\end{figure}

\begin{figure}
\includegraphics[scale=0.4]{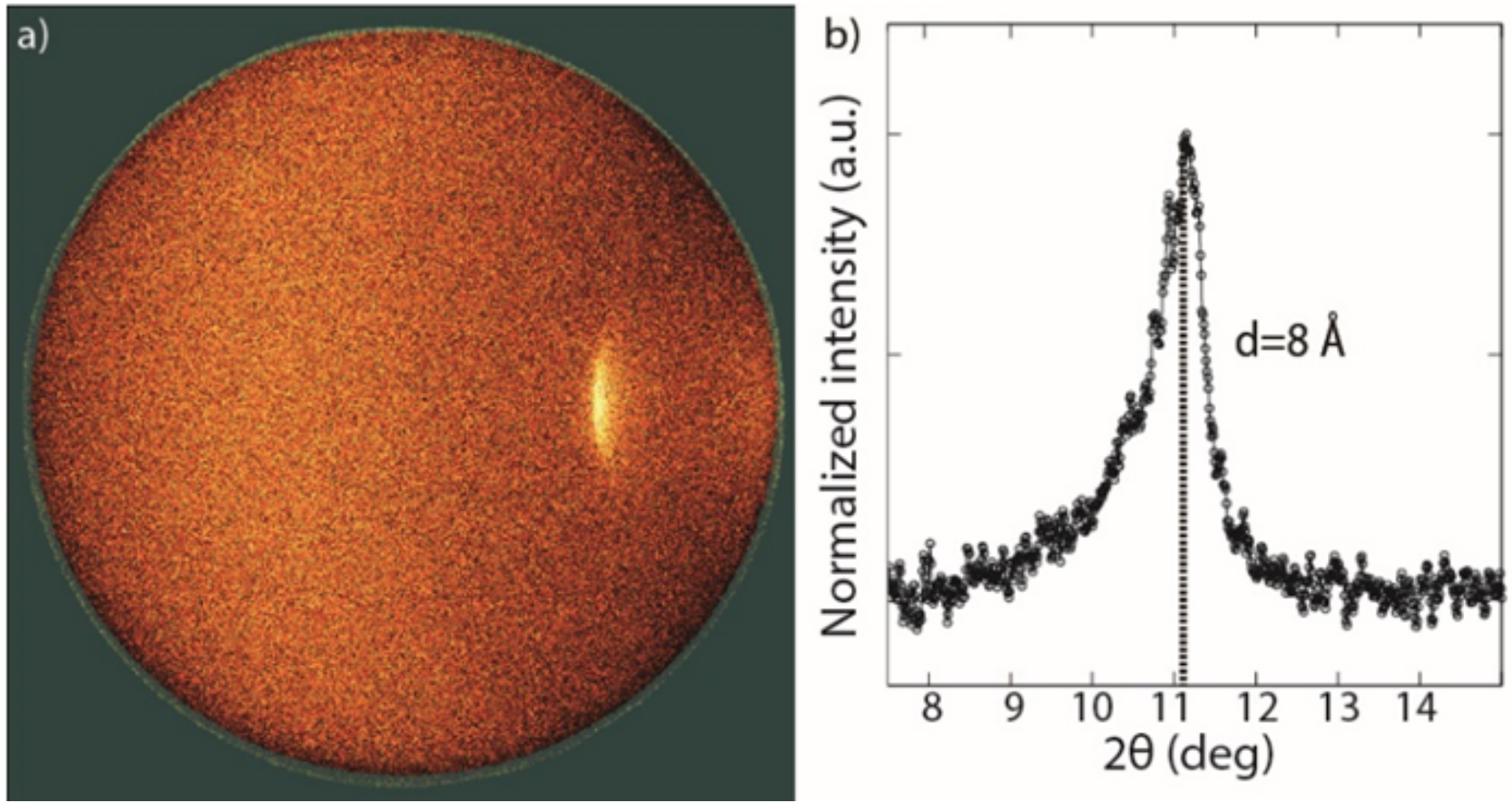}
\caption{\label{S3}(a) 2 d image of the Go membrane. (b)  XRD characterization of GO membrane showing a peak centered at $11.1^{\circ}$. }
\end{figure}

\begin{figure}
\includegraphics[scale=0.4]{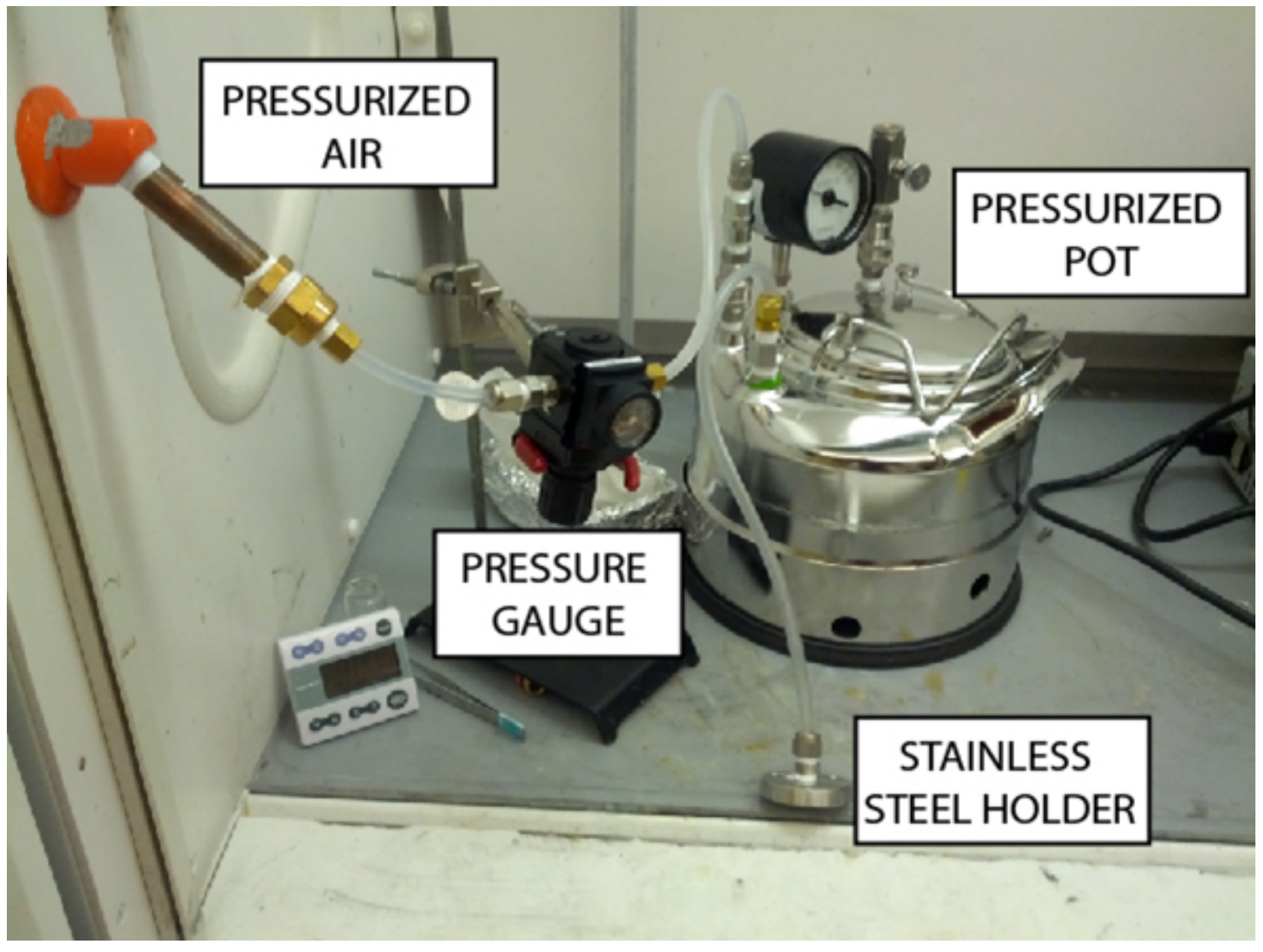}
\caption{\label{S4} Dead end filtration system.}
\end{figure}

\begin{figure}
\includegraphics[scale=1.0]{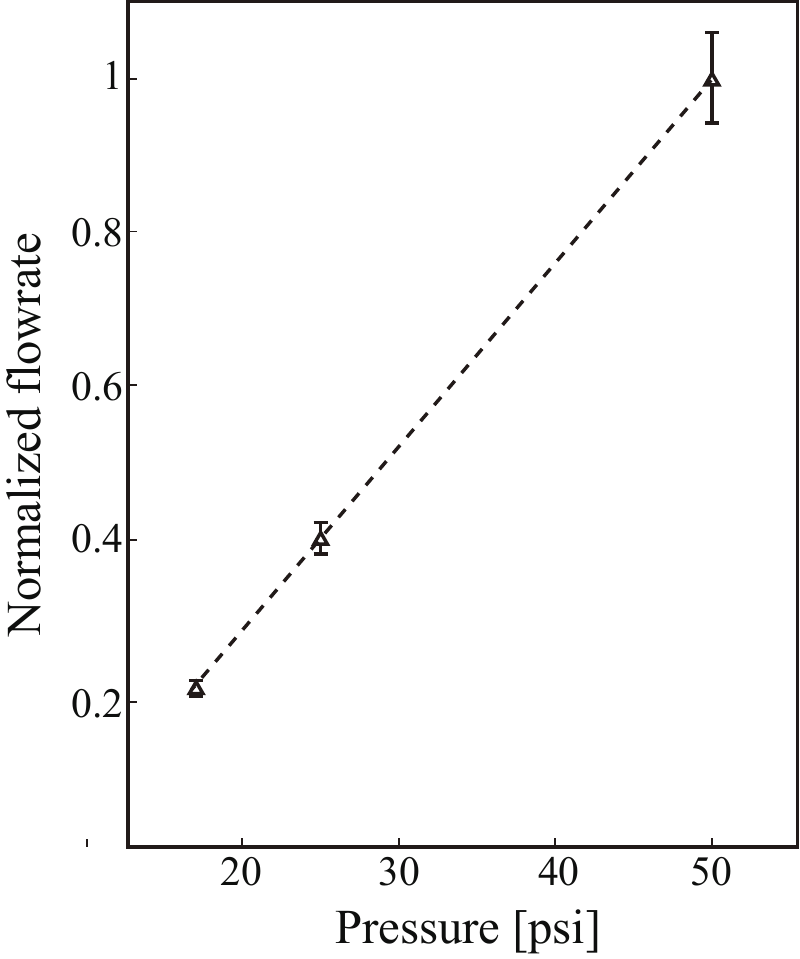}
\caption{\label{S5} Flowrate versus Pressure plot. The linear trend clearly denotes a Darcy behaviour of the GO membrane.}
\end{figure}

Figure S\;\ref{S2} represents SEM images obtained after depositing $2 nm$ $Pt/Pd$ layer onto the membranes via an EMS300R sputter coater. Figure S\;\ref{S2}a shows the bare PVDF membrane characterized by pores size of $\simeq 200 nm$. Figure S\;\ref{S2}b represents the PVDF membrane after the GO deposition via vacuum filtration. 
Figure S\;\ref{S2}c shows a cross section of the membrane, which highlights the extreme thinness of the GO layer (circa $30 nm$). It is also possible to observe that the GO layer conserves the morphology of the underneath PVDF membrane.
As explained in the main text, the XRD spectra was obtained by integrating a 2D diffraction pattern image (Fig. S\;\ref{S3}a). The XRD spectra (Fig. S\;\ref{S3}b) displays a distinct peak centered at $11.1^{\circ} $.  By using Bragg's law, this value can be converted to $0.8 nm$, which represents the spacing between the GO flakes.
The permeability tests were carried out with a custom-made dead-end filtration system (Figure S\;\ref{S4}). The filtration system includes a $3 L$ reservoir (e.g. pressure pot). The pressure was regulated with a pressure gauge from Ingersoll and the GO membrane was placed inside a stainless steel EDM Millipore filter holder.
Finally. fig. S\;\ref{S5} displays the normalized flowrate versus applied pressure for the GO membrane. As mentioned in the main text, the permeability is independent from the applied pressure and this is testified by the linear fitting.


\subsection{Some remarks on the slip length in nano-channel flows}
The possibility of slip flow was contemplated by Navier himself as early as in 1823, by postulating a 
direct proportionality between the slip velocity and the gradient of the velocity at the wall. For a one-dimensional flow across parallel plates at a distance $h$, the Navier slip-boundary adds a slip term $L_s/h$, on top of the parabolic Poiseuille bulk profile $1-(y/h)^2$. One may  therefore argue that the notion of slip-flow is 
perfectly compatible with continuum hydrodynamics. This is certainly true for the case of macroscopic flows, in 
which the slip term  $L_s/h\ll 1$. However, for nanoflows, where $h$ becomes comparable with $L_s$, this is no 
longer true, because the slip term  becomes dominant over the bulk one. Moreover, under such conditions, it is 
not obvious that the bulk component  still obeys a parabolic flow profile, because $L_s/h\sim 1$ means that non-
equilibrium is  as strong as equilibrium, thus contradicting the low-Knudsen assumption, $L_s/h\ll 1$,   which 
lies at the basis of the  continuum picture.

\begin{acknowledgments}
A.M., G.F. and S.S. were partially supported by the Integrated Mesoscale Architectures for Sustainable Catalysis (IMASC), an Energy Frontier Research Center funded by the US Department of Energy, Office of Science, Basic Energy Sciences under Award No. DE-SC0012573. C.A.A. would like to thank  Computational and Fluid Dynamics class (AP274) at Harvard, from where this work started.
This work made use of the Center for Nanoscale
Systems at Harvard University, a member of the National
Nanotechnology Infrastructure Network, supported (in part) by
the National Science Foundation under NSF award number
ECS-0335765.
\end{acknowledgments}
\bibliographystyle{apsrev4-1}

\begin{thebibliography}{37}%
\makeatletter
\providecommand \@ifxundefined [1]{%
 \@ifx{#1\undefined}
}%
\providecommand \@ifnum [1]{%
 \ifnum #1\expandafter \@firstoftwo
 \else \expandafter \@secondoftwo
 \fi
}%
\providecommand \@ifx [1]{%
 \ifx #1\expandafter \@firstoftwo
 \else \expandafter \@secondoftwo
 \fi
}%
\providecommand \natexlab [1]{#1}%
\providecommand \enquote  [1]{``#1''}%
\providecommand \bibnamefont  [1]{#1}%
\providecommand \bibfnamefont [1]{#1}%
\providecommand \citenamefont [1]{#1}%
\providecommand \href@noop [0]{\@secondoftwo}%
\providecommand \href [0]{\begingroup \@sanitize@url \@href}%
\providecommand \@href[1]{\@@startlink{#1}\@@href}%
\providecommand \@@href[1]{\endgroup#1\@@endlink}%
\providecommand \@sanitize@url [0]{\catcode `\\12\catcode `\$12\catcode
  `\&12\catcode `\#12\catcode `\^12\catcode `\_12\catcode `\%12\relax}%
\providecommand \@@startlink[1]{}%
\providecommand \@@endlink[0]{}%
\providecommand \url  [0]{\begingroup\@sanitize@url \@url }%
\providecommand \@url [1]{\endgroup\@href {#1}{\urlprefix }}%
\providecommand \urlprefix  [0]{URL }%
\providecommand \Eprint [0]{\href }%
\providecommand \doibase [0]{http://dx.doi.org/}%
\providecommand \selectlanguage [0]{\@gobble}%
\providecommand \bibinfo  [0]{\@secondoftwo}%
\providecommand \bibfield  [0]{\@secondoftwo}%
\providecommand \translation [1]{[#1]}%
\providecommand \BibitemOpen [0]{}%
\providecommand \bibitemStop [0]{}%
\providecommand \bibitemNoStop [0]{.\EOS\space}%
\providecommand \EOS [0]{\spacefactor3000\relax}%
\providecommand \BibitemShut  [1]{\csname bibitem#1\endcsname}%
\let\auto@bib@innerbib\@empty
\bibitem [{\citenamefont {Cohen-Tanugi}\ and\ \citenamefont
  {Grossman}(2012)}]{cohen2012water}%
  \BibitemOpen
  \bibfield  {author} {\bibinfo {author} {\bibfnamefont {D.}~\bibnamefont
  {Cohen-Tanugi}}\ and\ \bibinfo {author} {\bibfnamefont {J.~C.}\ \bibnamefont
  {Grossman}},\ }\href@noop {} {\bibfield  {journal} {\bibinfo  {journal} {Nano
  letters}\ }\textbf {\bibinfo {volume} {12}},\ \bibinfo {pages} {3602}
  (\bibinfo {year} {2012})}\BibitemShut {NoStop}%
\bibitem [{\citenamefont {Mishra}\ and\ \citenamefont
  {Ramaprabhu}(2011)}]{mishra2011functionalized}%
  \BibitemOpen
  \bibfield  {author} {\bibinfo {author} {\bibfnamefont {A.~K.}\ \bibnamefont
  {Mishra}}\ and\ \bibinfo {author} {\bibfnamefont {S.}~\bibnamefont
  {Ramaprabhu}},\ }\href@noop {} {\bibfield  {journal} {\bibinfo  {journal}
  {Desalination}\ }\textbf {\bibinfo {volume} {282}},\ \bibinfo {pages} {39}
  (\bibinfo {year} {2011})}\BibitemShut {NoStop}%
\bibitem [{\citenamefont {Julkapli}\ and\ \citenamefont
  {Bagheri}(2015)}]{julkapli2015graphene}%
  \BibitemOpen
  \bibfield  {author} {\bibinfo {author} {\bibfnamefont {N.~M.}\ \bibnamefont
  {Julkapli}}\ and\ \bibinfo {author} {\bibfnamefont {S.}~\bibnamefont
  {Bagheri}},\ }\href@noop {} {\bibfield  {journal} {\bibinfo  {journal}
  {International Journal of Hydrogen Energy}\ }\textbf {\bibinfo {volume}
  {40}},\ \bibinfo {pages} {948} (\bibinfo {year} {2015})}\BibitemShut
  {NoStop}%
\bibitem [{\citenamefont {Garberoglio}\ \emph {et~al.}(2007)\citenamefont
  {Garberoglio}, \citenamefont {Sega},\ and\ \citenamefont
  {Vallauri}}]{garberoglio2007inhomogeneity}%
  \BibitemOpen
  \bibfield  {author} {\bibinfo {author} {\bibfnamefont {G.}~\bibnamefont
  {Garberoglio}}, \bibinfo {author} {\bibfnamefont {M.}~\bibnamefont {Sega}}, \
  and\ \bibinfo {author} {\bibfnamefont {R.}~\bibnamefont {Vallauri}},\
  }\href@noop {} {\bibfield  {journal} {\bibinfo  {journal} {The Journal of
  chemical physics}\ }\textbf {\bibinfo {volume} {126}},\ \bibinfo {pages}
  {125103} (\bibinfo {year} {2007})}\BibitemShut {NoStop}%
\bibitem [{\citenamefont {Joseph}\ and\ \citenamefont
  {Aluru}(2008)}]{joseph2008carbon}%
  \BibitemOpen
  \bibfield  {author} {\bibinfo {author} {\bibfnamefont {S.}~\bibnamefont
  {Joseph}}\ and\ \bibinfo {author} {\bibfnamefont {N.}~\bibnamefont {Aluru}},\
  }\href@noop {} {\bibfield  {journal} {\bibinfo  {journal} {Nano letters}\
  }\textbf {\bibinfo {volume} {8}},\ \bibinfo {pages} {452} (\bibinfo {year}
  {2008})}\BibitemShut {NoStop}%
\bibitem [{\citenamefont {Holt}\ \emph {et~al.}(2006)\citenamefont {Holt},
  \citenamefont {Park}, \citenamefont {Wang}, \citenamefont {Stadermann},
  \citenamefont {Artyukhin}, \citenamefont {Grigoropoulos}, \citenamefont
  {Noy},\ and\ \citenamefont {Bakajin}}]{holt2006fast}%
  \BibitemOpen
  \bibfield  {author} {\bibinfo {author} {\bibfnamefont {J.~K.}\ \bibnamefont
  {Holt}}, \bibinfo {author} {\bibfnamefont {H.~G.}\ \bibnamefont {Park}},
  \bibinfo {author} {\bibfnamefont {Y.}~\bibnamefont {Wang}}, \bibinfo {author}
  {\bibfnamefont {M.}~\bibnamefont {Stadermann}}, \bibinfo {author}
  {\bibfnamefont {A.~B.}\ \bibnamefont {Artyukhin}}, \bibinfo {author}
  {\bibfnamefont {C.~P.}\ \bibnamefont {Grigoropoulos}}, \bibinfo {author}
  {\bibfnamefont {A.}~\bibnamefont {Noy}}, \ and\ \bibinfo {author}
  {\bibfnamefont {O.}~\bibnamefont {Bakajin}},\ }\href@noop {} {\bibfield
  {journal} {\bibinfo  {journal} {Science}\ }\textbf {\bibinfo {volume}
  {312}},\ \bibinfo {pages} {1034} (\bibinfo {year} {2006})}\BibitemShut
  {NoStop}%
\bibitem [{\citenamefont {Zuo}\ \emph {et~al.}(2009)\citenamefont {Zuo},
  \citenamefont {Shen}, \citenamefont {Ma},\ and\ \citenamefont
  {Guo}}]{zuo2009transport}%
  \BibitemOpen
  \bibfield  {author} {\bibinfo {author} {\bibfnamefont {G.}~\bibnamefont
  {Zuo}}, \bibinfo {author} {\bibfnamefont {R.}~\bibnamefont {Shen}}, \bibinfo
  {author} {\bibfnamefont {S.}~\bibnamefont {Ma}}, \ and\ \bibinfo {author}
  {\bibfnamefont {W.}~\bibnamefont {Guo}},\ }\href@noop {} {\bibfield
  {journal} {\bibinfo  {journal} {ACS nano}\ }\textbf {\bibinfo {volume} {4}},\
  \bibinfo {pages} {205} (\bibinfo {year} {2009})}\BibitemShut {NoStop}%
\bibitem [{\citenamefont {Majumder}\ \emph {et~al.}(2005)\citenamefont
  {Majumder}, \citenamefont {Chopra}, \citenamefont {Andrews},\ and\
  \citenamefont {Hinds}}]{majumder2005nanoscale}%
  \BibitemOpen
  \bibfield  {author} {\bibinfo {author} {\bibfnamefont {M.}~\bibnamefont
  {Majumder}}, \bibinfo {author} {\bibfnamefont {N.}~\bibnamefont {Chopra}},
  \bibinfo {author} {\bibfnamefont {R.}~\bibnamefont {Andrews}}, \ and\
  \bibinfo {author} {\bibfnamefont {B.~J.}\ \bibnamefont {Hinds}},\ }\href@noop
  {} {\bibfield  {journal} {\bibinfo  {journal} {Nature}\ }\textbf {\bibinfo
  {volume} {438}},\ \bibinfo {pages} {44} (\bibinfo {year} {2005})}\BibitemShut
  {NoStop}%
\bibitem [{\citenamefont {Nair}\ \emph {et~al.}(2012)\citenamefont {Nair},
  \citenamefont {Wu}, \citenamefont {Jayaram}, \citenamefont {Grigorieva},\
  and\ \citenamefont {Geim}}]{nair2012unimpeded}%
  \BibitemOpen
  \bibfield  {author} {\bibinfo {author} {\bibfnamefont {R.}~\bibnamefont
  {Nair}}, \bibinfo {author} {\bibfnamefont {H.}~\bibnamefont {Wu}}, \bibinfo
  {author} {\bibfnamefont {P.}~\bibnamefont {Jayaram}}, \bibinfo {author}
  {\bibfnamefont {I.}~\bibnamefont {Grigorieva}}, \ and\ \bibinfo {author}
  {\bibfnamefont {A.}~\bibnamefont {Geim}},\ }\href@noop {} {\bibfield
  {journal} {\bibinfo  {journal} {Science}\ }\textbf {\bibinfo {volume}
  {335}},\ \bibinfo {pages} {442} (\bibinfo {year} {2012})}\BibitemShut
  {NoStop}%
\bibitem [{\citenamefont {Hu}\ and\ \citenamefont {Mi}(2013)}]{hu2013enabling}%
  \BibitemOpen
  \bibfield  {author} {\bibinfo {author} {\bibfnamefont {M.}~\bibnamefont
  {Hu}}\ and\ \bibinfo {author} {\bibfnamefont {B.}~\bibnamefont {Mi}},\
  }\href@noop {} {\bibfield  {journal} {\bibinfo  {journal} {Environmental
  science \& technology}\ }\textbf {\bibinfo {volume} {47}},\ \bibinfo {pages}
  {3715} (\bibinfo {year} {2013})}\BibitemShut {NoStop}%
\bibitem [{\citenamefont {Jiang}\ \emph {et~al.}(2015)\citenamefont {Jiang},
  \citenamefont {Wang}, \citenamefont {Liu}, \citenamefont {Nie}, \citenamefont
  {Li}, \citenamefont {Wu}, \citenamefont {Zhang}, \citenamefont {Biswas},\
  and\ \citenamefont {Fortner}}]{jiang2015engineered}%
  \BibitemOpen
  \bibfield  {author} {\bibinfo {author} {\bibfnamefont {Y.}~\bibnamefont
  {Jiang}}, \bibinfo {author} {\bibfnamefont {W.-N.}\ \bibnamefont {Wang}},
  \bibinfo {author} {\bibfnamefont {D.}~\bibnamefont {Liu}}, \bibinfo {author}
  {\bibfnamefont {Y.}~\bibnamefont {Nie}}, \bibinfo {author} {\bibfnamefont
  {W.}~\bibnamefont {Li}}, \bibinfo {author} {\bibfnamefont {J.}~\bibnamefont
  {Wu}}, \bibinfo {author} {\bibfnamefont {F.}~\bibnamefont {Zhang}}, \bibinfo
  {author} {\bibfnamefont {P.}~\bibnamefont {Biswas}}, \ and\ \bibinfo {author}
  {\bibfnamefont {J.~D.}\ \bibnamefont {Fortner}},\ }\href@noop {} {\bibfield
  {journal} {\bibinfo  {journal} {Environmental science \& technology}\
  }\textbf {\bibinfo {volume} {49}},\ \bibinfo {pages} {6846} (\bibinfo {year}
  {2015})}\BibitemShut {NoStop}%
\bibitem [{\citenamefont {Han}\ \emph {et~al.}(2013)\citenamefont {Han},
  \citenamefont {Xu},\ and\ \citenamefont {Gao}}]{han2013ultrathin}%
  \BibitemOpen
  \bibfield  {author} {\bibinfo {author} {\bibfnamefont {Y.}~\bibnamefont
  {Han}}, \bibinfo {author} {\bibfnamefont {Z.}~\bibnamefont {Xu}}, \ and\
  \bibinfo {author} {\bibfnamefont {C.}~\bibnamefont {Gao}},\ }\href@noop {}
  {\bibfield  {journal} {\bibinfo  {journal} {Advanced Functional Materials}\
  }\textbf {\bibinfo {volume} {23}},\ \bibinfo {pages} {3693} (\bibinfo {year}
  {2013})}\BibitemShut {NoStop}%
\bibitem [{\citenamefont {Huang}\ \emph {et~al.}(2013)\citenamefont {Huang},
  \citenamefont {Song}, \citenamefont {Wei}, \citenamefont {Shi}, \citenamefont
  {Mao}, \citenamefont {Ying}, \citenamefont {Sun}, \citenamefont {Xu},\ and\
  \citenamefont {Peng}}]{huang2013ultrafast}%
  \BibitemOpen
  \bibfield  {author} {\bibinfo {author} {\bibfnamefont {H.}~\bibnamefont
  {Huang}}, \bibinfo {author} {\bibfnamefont {Z.}~\bibnamefont {Song}},
  \bibinfo {author} {\bibfnamefont {N.}~\bibnamefont {Wei}}, \bibinfo {author}
  {\bibfnamefont {L.}~\bibnamefont {Shi}}, \bibinfo {author} {\bibfnamefont
  {Y.}~\bibnamefont {Mao}}, \bibinfo {author} {\bibfnamefont {Y.}~\bibnamefont
  {Ying}}, \bibinfo {author} {\bibfnamefont {L.}~\bibnamefont {Sun}}, \bibinfo
  {author} {\bibfnamefont {Z.}~\bibnamefont {Xu}}, \ and\ \bibinfo {author}
  {\bibfnamefont {X.}~\bibnamefont {Peng}},\ }\href@noop {} {\bibfield
  {journal} {\bibinfo  {journal} {Nature communications}\ }\textbf {\bibinfo
  {volume} {4}} (\bibinfo {year} {2013})}\BibitemShut {NoStop}%
\bibitem [{\citenamefont {Sun}\ \emph {et~al.}(2016)\citenamefont {Sun},
  \citenamefont {Wang},\ and\ \citenamefont {Zhu}}]{sun2016recent}%
  \BibitemOpen
  \bibfield  {author} {\bibinfo {author} {\bibfnamefont {P.}~\bibnamefont
  {Sun}}, \bibinfo {author} {\bibfnamefont {K.}~\bibnamefont {Wang}}, \ and\
  \bibinfo {author} {\bibfnamefont {H.}~\bibnamefont {Zhu}},\ }\href@noop {}
  {\bibfield  {journal} {\bibinfo  {journal} {Advanced Materials}\ } (\bibinfo
  {year} {2016})}\BibitemShut {NoStop}%
\bibitem [{\citenamefont {Joshi}\ \emph {et~al.}(2014)\citenamefont {Joshi},
  \citenamefont {Carbone}, \citenamefont {Wang}, \citenamefont {Kravets},
  \citenamefont {Su}, \citenamefont {Grigorieva}, \citenamefont {Wu},
  \citenamefont {Geim},\ and\ \citenamefont {Nair}}]{joshi2014precise}%
  \BibitemOpen
  \bibfield  {author} {\bibinfo {author} {\bibfnamefont {R.}~\bibnamefont
  {Joshi}}, \bibinfo {author} {\bibfnamefont {P.}~\bibnamefont {Carbone}},
  \bibinfo {author} {\bibfnamefont {F.}~\bibnamefont {Wang}}, \bibinfo {author}
  {\bibfnamefont {V.}~\bibnamefont {Kravets}}, \bibinfo {author} {\bibfnamefont
  {Y.}~\bibnamefont {Su}}, \bibinfo {author} {\bibfnamefont {I.}~\bibnamefont
  {Grigorieva}}, \bibinfo {author} {\bibfnamefont {H.}~\bibnamefont {Wu}},
  \bibinfo {author} {\bibfnamefont {A.}~\bibnamefont {Geim}}, \ and\ \bibinfo
  {author} {\bibfnamefont {R.}~\bibnamefont {Nair}},\ }\href@noop {} {\bibfield
   {journal} {\bibinfo  {journal} {Science}\ }\textbf {\bibinfo {volume}
  {343}},\ \bibinfo {pages} {752} (\bibinfo {year} {2014})}\BibitemShut
  {NoStop}%
\bibitem [{\citenamefont {Wei}\ \emph {et~al.}(2014{\natexlab{a}})\citenamefont
  {Wei}, \citenamefont {Peng},\ and\ \citenamefont {Xu}}]{wei2014breakdown}%
  \BibitemOpen
  \bibfield  {author} {\bibinfo {author} {\bibfnamefont {N.}~\bibnamefont
  {Wei}}, \bibinfo {author} {\bibfnamefont {X.}~\bibnamefont {Peng}}, \ and\
  \bibinfo {author} {\bibfnamefont {Z.}~\bibnamefont {Xu}},\ }\href@noop {}
  {\bibfield  {journal} {\bibinfo  {journal} {Physical Review E}\ }\textbf
  {\bibinfo {volume} {89}},\ \bibinfo {pages} {012113} (\bibinfo {year}
  {2014}{\natexlab{a}})}\BibitemShut {NoStop}%
\bibitem [{\citenamefont {Wei}\ \emph {et~al.}(2014{\natexlab{b}})\citenamefont
  {Wei}, \citenamefont {Peng},\ and\ \citenamefont
  {Xu}}]{wei2014understanding}%
  \BibitemOpen
  \bibfield  {author} {\bibinfo {author} {\bibfnamefont {N.}~\bibnamefont
  {Wei}}, \bibinfo {author} {\bibfnamefont {X.}~\bibnamefont {Peng}}, \ and\
  \bibinfo {author} {\bibfnamefont {Z.}~\bibnamefont {Xu}},\ }\href@noop {}
  {\bibfield  {journal} {\bibinfo  {journal} {ACS applied materials \&
  interfaces}\ }\textbf {\bibinfo {volume} {6}},\ \bibinfo {pages} {5877}
  (\bibinfo {year} {2014}{\natexlab{b}})}\BibitemShut {NoStop}%
\bibitem [{\citenamefont {Montessori}\ \emph {et~al.}(2015)\citenamefont
  {Montessori}, \citenamefont {Prestininzi}, \citenamefont {La~Rocca},\ and\
  \citenamefont {Succi}}]{montessori2015lattice}%
  \BibitemOpen
  \bibfield  {author} {\bibinfo {author} {\bibfnamefont {A.}~\bibnamefont
  {Montessori}}, \bibinfo {author} {\bibfnamefont {P.}~\bibnamefont
  {Prestininzi}}, \bibinfo {author} {\bibfnamefont {M.}~\bibnamefont
  {La~Rocca}}, \ and\ \bibinfo {author} {\bibfnamefont {S.}~\bibnamefont
  {Succi}},\ }\href@noop {} {\bibfield  {journal} {\bibinfo  {journal}
  {Physical Review E}\ }\textbf {\bibinfo {volume} {92}},\ \bibinfo {pages}
  {043308} (\bibinfo {year} {2015})}\BibitemShut {NoStop}%
\bibitem [{\citenamefont {Kovtyukhova}\ \emph {et~al.}(1999)\citenamefont
  {Kovtyukhova}, \citenamefont {Ollivier}, \citenamefont {Martin},
  \citenamefont {Mallouk}, \citenamefont {Chizhik}, \citenamefont {Buzaneva},\
  and\ \citenamefont {Gorchinskiy}}]{kovtyukhova1999layer}%
  \BibitemOpen
  \bibfield  {author} {\bibinfo {author} {\bibfnamefont {N.~I.}\ \bibnamefont
  {Kovtyukhova}}, \bibinfo {author} {\bibfnamefont {P.~J.}\ \bibnamefont
  {Ollivier}}, \bibinfo {author} {\bibfnamefont {B.~R.}\ \bibnamefont
  {Martin}}, \bibinfo {author} {\bibfnamefont {T.~E.}\ \bibnamefont {Mallouk}},
  \bibinfo {author} {\bibfnamefont {S.~A.}\ \bibnamefont {Chizhik}}, \bibinfo
  {author} {\bibfnamefont {E.~V.}\ \bibnamefont {Buzaneva}}, \ and\ \bibinfo
  {author} {\bibfnamefont {A.~D.}\ \bibnamefont {Gorchinskiy}},\ }\href@noop {}
  {\bibfield  {journal} {\bibinfo  {journal} {Chemistry of Materials}\ }\textbf
  {\bibinfo {volume} {11}},\ \bibinfo {pages} {771} (\bibinfo {year}
  {1999})}\BibitemShut {NoStop}%
\bibitem [{\citenamefont {Qian}\ \emph {et~al.}(1992)\citenamefont {Qian},
  \citenamefont {d'Humi{\`e}res},\ and\ \citenamefont
  {Lallemand}}]{qian1992lattice}%
  \BibitemOpen
  \bibfield  {author} {\bibinfo {author} {\bibfnamefont {Y.}~\bibnamefont
  {Qian}}, \bibinfo {author} {\bibfnamefont {D.}~\bibnamefont
  {d'Humi{\`e}res}}, \ and\ \bibinfo {author} {\bibfnamefont {P.}~\bibnamefont
  {Lallemand}},\ }\href@noop {} {\bibfield  {journal} {\bibinfo  {journal} {EPL
  (Europhysics Letters)}\ }\textbf {\bibinfo {volume} {17}},\ \bibinfo {pages}
  {479} (\bibinfo {year} {1992})}\BibitemShut {NoStop}%
\bibitem [{\citenamefont {Higuera}\ \emph {et~al.}(1989)\citenamefont
  {Higuera}, \citenamefont {Succi},\ and\ \citenamefont
  {Benzi}}]{higuera1989lattice}%
  \BibitemOpen
  \bibfield  {author} {\bibinfo {author} {\bibfnamefont {F.}~\bibnamefont
  {Higuera}}, \bibinfo {author} {\bibfnamefont {S.}~\bibnamefont {Succi}}, \
  and\ \bibinfo {author} {\bibfnamefont {R.}~\bibnamefont {Benzi}},\
  }\href@noop {} {\bibfield  {journal} {\bibinfo  {journal} {EPL (Europhysics
  Letters)}\ }\textbf {\bibinfo {volume} {9}},\ \bibinfo {pages} {345}
  (\bibinfo {year} {1989})}\BibitemShut {NoStop}%
\bibitem [{\citenamefont {Succi}(2015)}]{succi2038lattice}%
  \BibitemOpen
  \bibfield  {author} {\bibinfo {author} {\bibfnamefont {S.}~\bibnamefont
  {Succi}},\ }\href@noop {} {\bibfield  {journal} {\bibinfo  {journal} {EPL
  (Europhysics Letters)}\ }\textbf {\bibinfo {volume} {109}},\ \bibinfo {pages}
  {50001} (\bibinfo {year} {2015})}\BibitemShut {NoStop}%
\bibitem [{\citenamefont {Succi}(2001)}]{succi2001lattice}%
  \BibitemOpen
  \bibfield  {author} {\bibinfo {author} {\bibfnamefont {S.}~\bibnamefont
  {Succi}},\ }\href@noop {} {\emph {\bibinfo {title} {The lattice Boltzmann
  equation: for fluid dynamics and beyond}}}\ (\bibinfo  {publisher} {Oxford
  university press},\ \bibinfo {year} {2001})\BibitemShut {NoStop}%
\bibitem [{\citenamefont {Wolf-Gladrow}(2000)}]{wolf2000lattice}%
  \BibitemOpen
  \bibfield  {author} {\bibinfo {author} {\bibfnamefont {D.~A.}\ \bibnamefont
  {Wolf-Gladrow}},\ }\href@noop {} {\emph {\bibinfo {title} {Lattice-gas
  cellular automata and lattice Boltzmann models: An Introduction}}}\ (\bibinfo
   {publisher} {Springer Science \& Business Media},\ \bibinfo {year}
  {2000})\BibitemShut {NoStop}%
\bibitem [{\citenamefont {Smiatek}\ \emph {et~al.}(2008)\citenamefont
  {Smiatek}, \citenamefont {Allen},\ and\ \citenamefont
  {Schmid}}]{smiatek2008tunable}%
  \BibitemOpen
  \bibfield  {author} {\bibinfo {author} {\bibfnamefont {J.}~\bibnamefont
  {Smiatek}}, \bibinfo {author} {\bibfnamefont {M.~P.}\ \bibnamefont {Allen}},
  \ and\ \bibinfo {author} {\bibfnamefont {F.}~\bibnamefont {Schmid}},\
  }\href@noop {} {\bibfield  {journal} {\bibinfo  {journal} {The European
  Physical Journal E}\ }\textbf {\bibinfo {volume} {26}},\ \bibinfo {pages}
  {115} (\bibinfo {year} {2008})}\BibitemShut {NoStop}%
\bibitem [{\citenamefont {Pastor}\ \emph {et~al.}(1988)\citenamefont {Pastor},
  \citenamefont {Brooks},\ and\ \citenamefont {Szabo}}]{pastor1988analysis}%
  \BibitemOpen
  \bibfield  {author} {\bibinfo {author} {\bibfnamefont {R.~W.}\ \bibnamefont
  {Pastor}}, \bibinfo {author} {\bibfnamefont {B.~R.}\ \bibnamefont {Brooks}},
  \ and\ \bibinfo {author} {\bibfnamefont {A.}~\bibnamefont {Szabo}},\
  }\href@noop {} {\bibfield  {journal} {\bibinfo  {journal} {Molecular
  Physics}\ }\textbf {\bibinfo {volume} {65}},\ \bibinfo {pages} {1409}
  (\bibinfo {year} {1988})}\BibitemShut {NoStop}%
\bibitem [{\citenamefont {Izaguirre}\ \emph {et~al.}(2001)\citenamefont
  {Izaguirre}, \citenamefont {Catarello}, \citenamefont {Wozniak},\ and\
  \citenamefont {Skeel}}]{izaguirre2001langevin}%
  \BibitemOpen
  \bibfield  {author} {\bibinfo {author} {\bibfnamefont {J.~A.}\ \bibnamefont
  {Izaguirre}}, \bibinfo {author} {\bibfnamefont {D.~P.}\ \bibnamefont
  {Catarello}}, \bibinfo {author} {\bibfnamefont {J.~M.}\ \bibnamefont
  {Wozniak}}, \ and\ \bibinfo {author} {\bibfnamefont {R.~D.}\ \bibnamefont
  {Skeel}},\ }\href@noop {} {\bibfield  {journal} {\bibinfo  {journal} {The
  Journal of chemical physics}\ }\textbf {\bibinfo {volume} {114}},\ \bibinfo
  {pages} {2090} (\bibinfo {year} {2001})}\BibitemShut {NoStop}%
\bibitem [{\citenamefont {Huang}\ \emph {et~al.}(2008)\citenamefont {Huang},
  \citenamefont {Sendner}, \citenamefont {Horinek}, \citenamefont {Netz},\ and\
  \citenamefont {Bocquet}}]{bocquet2008prl}%
  \BibitemOpen
  \bibfield  {author} {\bibinfo {author} {\bibfnamefont {D.~M.}\ \bibnamefont
  {Huang}}, \bibinfo {author} {\bibfnamefont {C.}~\bibnamefont {Sendner}},
  \bibinfo {author} {\bibfnamefont {D.}~\bibnamefont {Horinek}}, \bibinfo
  {author} {\bibfnamefont {R.~R.}\ \bibnamefont {Netz}}, \ and\ \bibinfo
  {author} {\bibfnamefont {L.}~\bibnamefont {Bocquet}},\ }\href {\doibase
  10.1103/PhysRevLett.101.226101} {\bibfield  {journal} {\bibinfo  {journal}
  {Phys. Rev. Lett.}\ }\textbf {\bibinfo {volume} {101}},\ \bibinfo {pages}
  {226101} (\bibinfo {year} {2008})}\BibitemShut {NoStop}%
\bibitem [{\citenamefont {Sega}\ \emph {et~al.}(2013)\citenamefont {Sega},
  \citenamefont {Sbragaglia}, \citenamefont {Biferale},\ and\ \citenamefont
  {Succi}}]{sega2013regularization}%
  \BibitemOpen
  \bibfield  {author} {\bibinfo {author} {\bibfnamefont {M.}~\bibnamefont
  {Sega}}, \bibinfo {author} {\bibfnamefont {M.}~\bibnamefont {Sbragaglia}},
  \bibinfo {author} {\bibfnamefont {L.}~\bibnamefont {Biferale}}, \ and\
  \bibinfo {author} {\bibfnamefont {S.}~\bibnamefont {Succi}},\ }\href@noop {}
  {\bibfield  {journal} {\bibinfo  {journal} {Soft Matter}\ }\textbf {\bibinfo
  {volume} {9}},\ \bibinfo {pages} {8526} (\bibinfo {year} {2013})}\BibitemShut
  {NoStop}%
\bibitem [{\citenamefont {Horbach}\ and\ \citenamefont
  {Succi}(2006)}]{horbach2006lattice}%
  \BibitemOpen
  \bibfield  {author} {\bibinfo {author} {\bibfnamefont {J.}~\bibnamefont
  {Horbach}}\ and\ \bibinfo {author} {\bibfnamefont {S.}~\bibnamefont
  {Succi}},\ }\href@noop {} {\bibfield  {journal} {\bibinfo  {journal}
  {Physical review letters}\ }\textbf {\bibinfo {volume} {96}},\ \bibinfo
  {pages} {224503} (\bibinfo {year} {2006})}\BibitemShut {NoStop}%
\bibitem [{\citenamefont {Fyta}\ \emph {et~al.}(2006)\citenamefont {Fyta},
  \citenamefont {Melchionna}, \citenamefont {Kaxiras},\ and\ \citenamefont
  {Succi}}]{fyta2006multiscale}%
  \BibitemOpen
  \bibfield  {author} {\bibinfo {author} {\bibfnamefont {M.~G.}\ \bibnamefont
  {Fyta}}, \bibinfo {author} {\bibfnamefont {S.}~\bibnamefont {Melchionna}},
  \bibinfo {author} {\bibfnamefont {E.}~\bibnamefont {Kaxiras}}, \ and\
  \bibinfo {author} {\bibfnamefont {S.}~\bibnamefont {Succi}},\ }\href@noop {}
  {\bibfield  {journal} {\bibinfo  {journal} {Multiscale Modeling \&
  Simulation}\ }\textbf {\bibinfo {volume} {5}},\ \bibinfo {pages} {1156}
  (\bibinfo {year} {2006})}\BibitemShut {NoStop}%
\bibitem [{\citenamefont {Muscatello}\ \emph {et~al.}(2016)\citenamefont
  {Muscatello}, \citenamefont {Jaeger}, \citenamefont {Matar},\ and\
  \citenamefont {M\"uller}}]{muscatello2016opt}%
  \BibitemOpen
  \bibfield  {author} {\bibinfo {author} {\bibfnamefont {J.}~\bibnamefont
  {Muscatello}}, \bibinfo {author} {\bibfnamefont {F.}~\bibnamefont {Jaeger}},
  \bibinfo {author} {\bibfnamefont {O.~K.}\ \bibnamefont {Matar}}, \ and\
  \bibinfo {author} {\bibfnamefont {E.~A.}\ \bibnamefont {M\"uller}},\ }\href
  {\doibase 10.1021/acsami.5b12112} {\bibfield  {journal} {\bibinfo  {journal}
  {ACS Applied Materials \& Interfaces}\ }\textbf {\bibinfo {volume} {8}},\
  \bibinfo {pages} {12330} (\bibinfo {year} {2016})},\ \bibinfo {note} {pMID:
  27121070},\ \Eprint
  {http://arxiv.org/abs/http://dx.doi.org/10.1021/acsami.5b12112}
  {http://dx.doi.org/10.1021/acsami.5b12112} \BibitemShut {NoStop}%
\bibitem [{\citenamefont {Amadei}\ \emph {et~al.}(2016)\citenamefont {Amadei},
  \citenamefont {Stein}, \citenamefont {Silverberg}, \citenamefont {Wardle},\
  and\ \citenamefont {Vecitis}}]{amadei2016fabrication}%
  \BibitemOpen
  \bibfield  {author} {\bibinfo {author} {\bibfnamefont {C.~A.}\ \bibnamefont
  {Amadei}}, \bibinfo {author} {\bibfnamefont {I.~Y.}\ \bibnamefont {Stein}},
  \bibinfo {author} {\bibfnamefont {G.~J.}\ \bibnamefont {Silverberg}},
  \bibinfo {author} {\bibfnamefont {B.}~\bibnamefont {Wardle}}, \ and\ \bibinfo
  {author} {\bibfnamefont {C.}~\bibnamefont {Vecitis}},\ }\href@noop {}
  {\bibfield  {journal} {\bibinfo  {journal} {Nanoscale}\ } (\bibinfo {year}
  {2016})}\BibitemShut {NoStop}%
\bibitem [{\citenamefont {Xia}\ \emph {et~al.}(2015)\citenamefont {Xia},
  \citenamefont {Ni}, \citenamefont {Zhu}, \citenamefont {Zhao},\ and\
  \citenamefont {Li}}]{xia2015ultrathin}%
  \BibitemOpen
  \bibfield  {author} {\bibinfo {author} {\bibfnamefont {S.}~\bibnamefont
  {Xia}}, \bibinfo {author} {\bibfnamefont {M.}~\bibnamefont {Ni}}, \bibinfo
  {author} {\bibfnamefont {T.}~\bibnamefont {Zhu}}, \bibinfo {author}
  {\bibfnamefont {Y.}~\bibnamefont {Zhao}}, \ and\ \bibinfo {author}
  {\bibfnamefont {N.}~\bibnamefont {Li}},\ }\href@noop {} {\bibfield  {journal}
  {\bibinfo  {journal} {Desalination}\ }\textbf {\bibinfo {volume} {371}},\
  \bibinfo {pages} {78} (\bibinfo {year} {2015})}\BibitemShut {NoStop}%
\bibitem [{\citenamefont {Ying}\ \emph {et~al.}(2014)\citenamefont {Ying},
  \citenamefont {Sun}, \citenamefont {Wang}, \citenamefont {Fan},\ and\
  \citenamefont {Peng}}]{ying2014plane}%
  \BibitemOpen
  \bibfield  {author} {\bibinfo {author} {\bibfnamefont {Y.}~\bibnamefont
  {Ying}}, \bibinfo {author} {\bibfnamefont {L.}~\bibnamefont {Sun}}, \bibinfo
  {author} {\bibfnamefont {Q.}~\bibnamefont {Wang}}, \bibinfo {author}
  {\bibfnamefont {Z.}~\bibnamefont {Fan}}, \ and\ \bibinfo {author}
  {\bibfnamefont {X.}~\bibnamefont {Peng}},\ }\href@noop {} {\bibfield
  {journal} {\bibinfo  {journal} {RSC Advances}\ }\textbf {\bibinfo {volume}
  {4}},\ \bibinfo {pages} {21425} (\bibinfo {year} {2014})}\BibitemShut
  {NoStop}%
\bibitem [{\citenamefont {Goh}\ \emph {et~al.}(2015)\citenamefont {Goh},
  \citenamefont {Jiang}, \citenamefont {Karahan}, \citenamefont {Zhai},
  \citenamefont {Wei}, \citenamefont {Yu}, \citenamefont {Fane}, \citenamefont
  {Wang},\ and\ \citenamefont {Chen}}]{goh2015all}%
  \BibitemOpen
  \bibfield  {author} {\bibinfo {author} {\bibfnamefont {K.}~\bibnamefont
  {Goh}}, \bibinfo {author} {\bibfnamefont {W.}~\bibnamefont {Jiang}}, \bibinfo
  {author} {\bibfnamefont {H.~E.}\ \bibnamefont {Karahan}}, \bibinfo {author}
  {\bibfnamefont {S.}~\bibnamefont {Zhai}}, \bibinfo {author} {\bibfnamefont
  {L.}~\bibnamefont {Wei}}, \bibinfo {author} {\bibfnamefont {D.}~\bibnamefont
  {Yu}}, \bibinfo {author} {\bibfnamefont {A.~G.}\ \bibnamefont {Fane}},
  \bibinfo {author} {\bibfnamefont {R.}~\bibnamefont {Wang}}, \ and\ \bibinfo
  {author} {\bibfnamefont {Y.}~\bibnamefont {Chen}},\ }\href@noop {} {\bibfield
   {journal} {\bibinfo  {journal} {Advanced Functional Materials}\ }\textbf
  {\bibinfo {volume} {25}},\ \bibinfo {pages} {7348} (\bibinfo {year}
  {2015})}\BibitemShut {NoStop}%
\bibitem [{\citenamefont {Haynes}(2014)}]{haynes2014crc}%
  \BibitemOpen
  \bibfield  {author} {\bibinfo {author} {\bibfnamefont {W.~M.}\ \bibnamefont
  {Haynes}},\ }\href@noop {} {\emph {\bibinfo {title} {CRC handbook of
  chemistry and physics}}}\ (\bibinfo  {publisher} {CRC press},\ \bibinfo
  {year} {2014})\BibitemShut {NoStop}%
\end{thebibliography}%
%
\end{document}